\documentclass[twocolumn]{aastex63}

\usepackage{color}
\usepackage{graphicx}
\usepackage{amsmath}
\usepackage{amssymb}
\hypersetup{pdfauthor={Name}}

\def\ra#1#2#3{#1$^{\rm h}$#2$^{\rm m}$#3$^{\rm s}$}
\def\dec#1#2#3{$#1^\circ#2'#3''$}
\def\grb{GRB\,200522A}
\def\nod{\nodata}
\def\swift{{\it Swift}}
\newcommand{\EK}{\ensuremath{E_{\rm K}}}
\newcommand{\EKiso}{\ensuremath{E_{\rm K,iso}}}

\newcommand{\epse}{\ensuremath{\epsilon_{\rm e}}}
\newcommand{\epsb}{\ensuremath{\epsilon_{\rm B}}}
\newcommand{\dens}{\ensuremath{n_{0}}}
\newcommand{\tjet}{\ensuremath{t_{\rm jet}}}
\newcommand{\thetajet}{\ensuremath{\theta_{\rm jet}}}
\newcommand{\nua}{\ensuremath{\nu_{\rm a}}}
\newcommand{\numax}{\ensuremath{\nu_{\rm m}}}
\newcommand{\nuc}{\ensuremath{\nu_{\rm c}}}
\newcommand{\nuX}{\ensuremath{\nu_{\rm X}}}

\newcommand{\rp}{\emph{r}-process}
\newcommand{\ye}{\ensuremath{Y_{\rm e}}}
\newcommand{\dote}{\ensuremath{\dot{\epsilon}_{\rm rp}}}
\newcommand{\doteo}{\ensuremath{\dot{\epsilon}_{\rm rp,0}}}

\graphicspath{{./}{figures/}}

\shorttitle{GRB 200522A}
\shortauthors{Fong et al.}

\begin{document}
\sloppy

\title{The Broad-band Counterpart of the Short GRB 200522A at $z=0.5536$: A Luminous Kilonova or a Collimated Outflow with a Reverse Shock?}

\correspondingauthor{W. Fong}
\email{wfong@northwestern.edu}

\newcommand{\NU}{\affiliation{Center for Interdisciplinary Exploration and Research in Astrophysics (CIERA) and Department of Physics and Astronomy, Northwestern University, Evanston, IL 60208, USA}}
\newcommand{\CfA}{\affiliation{Center for Astrophysics\:$|$\:Harvard \& Smithsonian, 60 Garden St. Cambridge, MA 02138, USA}}
\newcommand{\OU}{\affiliation{Astrophysical Institute, Department of Physics and Astronomy, 251B Clippinger Lab, Ohio University, Athens, OH 45701, USA}}
\newcommand{\Adler}{\affiliation{The Adler Planetarium, Chicago, IL 60605, USA}}
\newcommand{\GeminiN}{\affiliation{Gemini Observatory/NSF's NOIRLab, 670 N. A'ohoku Place, Hilo, HI, 96720, USA}}
\newcommand{\GSFC}{\affiliation{Astroparticle Physics Laboratory, NASA Goddard Space Flight Center, Mail Code 661, Greenbelt, MD 20771, USA}}
\newcommand{\UMD}{\affiliation{Joint Space-Science Institute, University of Maryland, College Park, MD 20742, USA}}
\newcommand{\GWU}{\affiliation{Department of Physics, The George Washington University, Washington, DC 20052, USA}}
\newcommand{\Leicester}{\affiliation{School of Physics and Astronomy, University of Leicester, University Road, Leicester, LE1 7RH, UK}}
\newcommand{\Marin}{\affiliation{College of Marin, 120 Kent Avenue, Kentfield 94904 CA, USA}}
\newcommand{\UVI}{\affiliation{University of the Virgin Islands, \#2 Brewers Bay Road, Charlotte Amalie, 00802 USVI, USA}}
\newcommand{\Radboud}{\affiliation{Department of Astrophysics/IMAPP, Radboud University, 6525 AJ Nijmegen, The Netherlands}}
\newcommand{\Warwick}{\affiliation{Department of Physics, University of Warwick, Coventry, CV4 7AL, UK}}
\newcommand{\Birmingham}{\affiliation{Birmingham Institute for Gravitational Wave Astronomy and School of Physics and Astronomy, University of Birmingham, Birmingham B15 2TT, UK}}
\newcommand{\Edinburgh}{\affiliation{Institute for Astronomy, University of Edinburgh, Royal Observatory, Blackford Hill, EH9 3HJ, UK}}
\newcommand{\Caltech}{\affiliation{Cahill Center for Astrophysics, California Institute of Technology, 1200 E. California Blvd. Pasadena, CA 91125, USA}}
\newcommand{\Bath}{\affiliation{Department of Physics, University of Bath, Claverton Down, Bath, BA2 7AY, UK}}
\newcommand{\Einstein}{\altaffiliation{NASA Einstein Fellow}}
\newcommand{\LJMU}{\affiliation{Astrophysics Research Institute, Liverpool John Moores University, 146 Brownlow Hill, Liverpool L3 5RF, UK}}
\newcommand{\Purdue}{\affiliation{Purdue University, 
Department of Physics and Astronomy, 525 Northwestern Avenue, West Lafayette, IN 47907, USA}}
\newcommand{\UCSC}{\affiliation{Department of Astronomy and Astrophysics, University of California, Santa Cruz, CA 95064, USA}}
\newcommand{\UCB}{\affiliation{Astronomy Department and Theoretical Astrophysics Center, University of California, Berkeley, Berkeley, CA 94720, USA}}
\newcommand{\CCA}{\affiliation{Center for Computational Astrophysics, Flatiron Institute, 162 W. 5th Avenue, New York, NY 10011, USA}}
\newcommand{\Columbia}{\affiliation{Department of Physics and Columbia Astrophysics Laboratory, Columbia University, New York, NY 10027, USA}}
\newcommand{\LANL}{\affiliation{Center for Theoretical Astrophysics, Los Alamos National Laboratory, Los Alamos, NM, 87545, USA}}

\author[0000-0002-7374-935X]{W. Fong}
\NU

\author[0000-0003-1792-2338]{T. Laskar}
\Bath

\author[0000-0002-9267-6213]{J. Rastinejad} 
\NU

\author[0000-0003-3937-0618]{A. Rouco Escorial} 
\NU

\author[0000-0001-9915-8147]{G.~Schroeder}
\NU

\author[0000-0003-3340-4784]{J. Barnes}
\Einstein\Columbia

\author[0000-0002-5740-7747]{C. D. Kilpatrick} 
\NU\UCSC

\author[0000-0001-8340-3486]{K. Paterson}
\NU

\author[0000-0002-9392-9681]{E. Berger} 
\CfA

\author[0000-0002-4670-7509]{B.~D.~Metzger}
\CCA\Columbia

\author[0000-0002-9363-8606]{Y. Dong}
\NU\Purdue

\author[0000-0002-2028-9329]{A. E. Nugent}
\NU

\author[0000-0001-6548-3777]{R. Strausbaugh} 
\UVI

\author[0000-0003-0526-2248]{P. K. Blanchard}
\NU

\author[0000-0001-9652-8384]{A.~Goyal}
\NU

\author{A. Cucchiara} 
\Marin\UVI

\author[0000-0003-0794-5982]{G. Terreran}
\NU

\author[0000-0002-8297-2473]{K. D. Alexander}
\Einstein\NU

\author[0000-0003-0307-9984]{T.~Eftekhari}
\CfA

\author{C.~Fryer}
\LANL

\author[0000-0001-8405-2649]{B.~Margalit}
\Einstein\UCB

\author[0000-0002-8297-2473]{R. Margutti}
\NU

\author[0000-0002-2555-3192]{M.~Nicholl}
\Birmingham\Edinburgh

\begin{abstract}
We present the discovery of the radio afterglow and near-infrared (NIR) counterpart of the {\it Swift} short GRB\,200522A, located at a small projected offset of $\approx 1$~kpc from the center of a young, star-forming host galaxy at $z=0.5536$. The radio and X-ray luminosities of the afterglow are consistent with those of on-axis cosmological short GRBs. The NIR counterpart, revealed by our {\it HST} observations at a rest-frame time of $\approx2.3$~days, has a luminosity of $\approx (1.3-1.7) \times 10^{42}$~erg~s$^{-1}$. This is substantially lower than on-axis short GRB afterglow detections, but is a factor of $\approx 8$--$17$ more luminous than the kilonova of GW170817, and significantly more luminous than any kilonova candidate for which comparable observations exist. The combination of the counterpart's color ($i-y = -0.08\pm 0.21$; rest-frame) and luminosity cannot be explained by standard radioactive heating alone. We present two scenarios to interpret the broad-band behavior of \grb: a synchrotron forward shock with a luminous kilonova (potentially boosted by magnetar energy deposition), or forward and reverse shocks from a $\approx14^{\circ}$, relativistic ($\Gamma_0\gtrsim80$) jet. Models which include a combination of enhanced radioactive heating rates, low-lanthanide mass fractions, or additional sources of heating from late-time central engine activity may provide viable alternate explanations. If a stable magnetar was indeed produced in \grb, we predict that late-time radio emission will be detectable starting $\approx 0.3$--$6$~years after the burst for a deposited energy of $\approx 10^{53}$~erg. Counterparts of similar luminosity to \grb\ associated with gravitational wave events will be detectable with current optical searches to $\approx\!250$~Mpc.
\end{abstract}


\section{Introduction}
\label{sec:intro}

Short-duration $\gamma$-ray bursts (GRBs) are energetic explosions with isotropic energy scales of order $\sim 10^{51}$~erg, and are detected to $z \approx 2$ \citep{npp92,gbb+08,ber14,lsb+16,pfn+20}. They have prompt $\gamma$-ray emission ($T_{\rm 90}<2$~s; \citealt{kmf+93,nak07}) and broad-band, synchrotron afterglow emission at radio to X-ray wavelengths as a result of collimated, relativistic material interacting with the circumburst environment \citep{sp95,mr97}. In the context of their likely binary neutron star (BNS) merger progenitors \citep{ber14,gw170817grb}, the non-thermal afterglows of short GRBs are expected to be accompanied by a thermal $r$-process kilonova \citep{lp98,mmd+10} powered by the radioactive decay of neutron-rich material synthesized in the merger. For short GRBs where the collimated outflow is viewed on-axis, the afterglow is expected to outshine the kilonova emission at optical wavelengths on $\lesssim 1$~day timescales. On $\gtrsim 1$~day timescales, the kilonova emission may dominate the observed optical and near-infrared (NIR) light, depending on the precise explosion properties of the afterglow (e.g., the kinetic energy, jet geometry) and the circumburst medium, as well as the mass, composition, and geometry of the kilonova ejecta (e.g., \citealt{bk13,wkf+17,met19}). Indeed, the four kilonova candidates associated with short GRBs have all been detected on timescales of $\gtrsim 1$~day \citep{bfc13,tlf+13,jhl+16,trp+18,ltl+19,tcb+19}. The optical and NIR emission of short GRBs and BNS mergers is thus a complex interplay between the non-thermal (potentially) jetted synchrotron emission and the thermal kilonova which results from heavy element nucleosynthesis.

In general, the radio band is observationally more straightforward for short GRBs, as the primary expected emission component is from the afterglow forward shock. However, despite routine, rapid follow-up observations, only seven short GRBs discovered by the Neil Gehrels {\it Swift} Observatory ({\it Swift}; \citealt{ggg+04}) have detected radio afterglows \citep{fbm+15}, or $\approx\!5\%$ of the entire {\it Swift} short GRB sample \citep{lsb+16}. Rapid-response, radio observations at $\lesssim 1$~day have enabled the detection of early excess emission compared to expectations from the forward shock model, interpreted as reverse shock emission in two events, GRBs\,051221A and 160821B \citep{sbk+06,llo18,ltl+19}. As a population, the {\it lack} of optical and radio afterglow emission for a majority of short GRBs is a direct reflection of their low beaming-corrected kinetic energy scales ($\approx 10^{49}$~erg, two orders of magnitude lower than long-duration GRBs; \citealt{pan06,gbb+08}), and their low circumburst densities of $\approx 10^{-3}-10^{-2}$~cm$^{-3}$ \citep{pkn01,sbk+06,fbm+15,obk20}.

Short GRBs also exhibit an extended spatial distribution with respect to their host galaxies, as well as to their host light distributions \citep{ber10,fbf10,fb13,tlt+14}. Their hosts have a range of stellar population ages of $\approx 0.5-8$~Gyr \citep{lb10,nfd+20}, which can naturally be explained by the wide expected range of delay times for their BNS merger progenitors \citep{bpb+06,pfn+20}. The low densities, weak correlation with host stellar mass or star formation, and origin from a diverse range of host galaxies are all hallmarks of the short GRB population \citep{zr07,fb13,fbc+13,tlt+14,wfs+18,obk20}.

The detection of kilonovae associated with short GRBs has been challenging, due to a combination of the faint expected emission and cosmological distances, making sufficient follow-up observations difficult with current resources. The four kilonova candidates associated with short GRBs, as well as the kilonova associated with the BNS merger GW170817, have luminosities and colors that can be explained by standard radioactive heating \citep{bkw+16,kmb+17}. The kilonova of GW170817 has a well-sampled multi-band light curve \citep{Andreoni17,Arcavi17_2,cbv+17,cbk+17,cfk+17,Diaz17,Drout17,Kasliwal17,Lipunov17,nbk+17,Pian17,Pozanenko17,scj+17,Tanvir17,Troja17,Utsumi17,vsy+17,vgb+17}, providing a benchmark for radioactively-powered kilonovae. The remaining short GRB-kilonova candidates are more sparsely-sampled and have been detected in a variety of rest-frame bands (optical and NIR), but overall exhibit an evolution from blue to redder colors with time. In addition the range of observed luminosities for the majority of events are $\approx (1-5) \times 10^{41}$~erg~s$^{-1}$. If all are in fact kilonovae, this demonstrates the diversity of kilonova emission resulting from BNS mergers (e.g., \citealt{acd+19,glt+18,rsm+20}). However, if the short GRB progenitor produces a hypermassive or supramassive neutron star that is at least temporarily stable to collapse, or even an indefinitely stable remnant, a combination of disk winds, neutrino irradiation, and spin-down energy may also be imprinted on the kilonova signal or X-ray emission, resulting in even larger luminosities and bluer colors \citep{mf14,mp14,kfm15,met19}. Variations on the ejecta morphology or sources of heating, independent of the presence of a stable remnant, may have similar effects \citep{kit15,rfk+17,bkw+16,met19,kwf20}. Thus far, there has not been a clear case of an observed kilonova or kilonova candidate which required the existence of a stable neutron star remnant, or major modifications to standard kilonova models.

Here, we present X-ray, optical, NIR and radio observations of the short \grb\ and its star-forming host galaxy at $z=0.5536$. These observations reveal an unusual broad-band counterpart that is not easily explained by a single emission component. In Section~\ref{sec:obs} we present the {\it Swift} burst discovery, the discovery of the radio and NIR counterparts with the VLA and {\it HST}, and observations of the host galaxy with Keck and archival data. In Sections~\ref{sec:bb_nirexcess} and \ref{sec:bb_alternative}, we introduce two scenarios to explain the peculiar broad-band behavior of \grb: a forward shock with a NIR excess, or a combination of forward and reverse shocks with a wide-angle jet. We present our host galaxy modeling, and derived stellar population, morphological, and local properties in Section~\ref{sec:host}. In Section~\ref{sec:kn} we introduce radioactively-powered and magnetar-boosted kilonova models to explain the NIR excess emission of \grb, and compare the NIR luminosity to the landscape of known or candidate kilonovae. In Section~\ref{sec:discussion}, we compare \grb\ to the population of short GRBs in terms of its transient and host galaxy properties, introduce a radio catalog of short GRB afterglow detections, and discuss implications for detectability. Finally, we conclude and offer a future outlook in Section~\ref{sec:conc}.

Unless otherwise stated, all observations are reported in AB mag and have been corrected for Galactic extinction in the direction of the burst of $A_{\rm V}=0.07$~mag \citep{sf11}. We employ a standard cosmology of $H_{0}$ = 69.6~km~s$^{-1}$~Mpc$^{-1}$, $\Omega_{M}$ = 0.286, $\Omega_{vac}$ = 0.714 \citep{blw+14}.

\section{Observations \& Data Analysis}
\label{sec:obs}

\subsection{Burst discovery}
\label{sec:disc}

\grb\ was discovered by the Burst Alert Telescope (BAT) on-board \swift\ \citep{ggg+04} on 2020 May 22 at 11:41:34 UT \citep{gcn27778}. The BAT position was refined to RA=\ra{00}{22}{40.3}, Dec=$-$\dec{00}{15}{49.9} (J2000) with an uncertainty of $1.59'$ in radius (90\% confidence; \citealt{gcn27793}). The \swift\ X-ray Telescope (XRT) began observations of the field of \grb\ at $\delta t=83.4$~s (where $\delta t$ is defined as the time since the BAT trigger) and detected an uncatalogued X-ray source within the BAT position, later refined to an enhanced position of RA=\ra{00}{22}{43.68}, Dec=$-$\dec{00}{16}{59.4} with a $2.2''$-radius positional uncertainty ($90\%$ confidence; \citealt{gcn27780,gtb+07,ebp+09}). The duration of the burst, with $T_{90}=0.62 \pm 0.08$~s (15-350~keV), combined with the hardness ratio of 1.46 (fluence ratio, $S(50-100)$~keV/$S(25-50)$~keV) place \grb\ solidly in the category of short, hard GRBs \citep{lsb+16}. We measure a \swift/BAT fluence of $S_{\gamma}=(1.04\pm0.14) \times 10^{-7}$~erg~cm$^{-2}$ ($15-150$~keV, 90\% confidence), consistent with the results of \citet{gcn27793}.

Upon a detailed inspection of the \grb\ 64-ms BAT light curve, we find a multi-peaked structure in the main GRB pulse. We also note a precursor signal prior to the main pulse between $\delta t=-0.35$~s and $\delta t=-0.25$~s. Constructing an image over this time interval in the $25-100$~keV band, we derive a source significance for the precursor of $3.9\sigma$. The spectrum of the precursor signal is poorly constrained, but is consistent with a hard spectrum characterized by photon index, $\Gamma_{\gamma}=0.86 \pm 0.70$. For \grb, the power-law (PL) and cut-off power-law (CPL) models provide comparable fits to the T$_{100}$ spectrum. Here, we employ the CPL model since it provides a constraint on the break energy of the spectrum, and therefore a more accurate estimate of the integrated energy. We obtain the best-fit values of $\Gamma_{\gamma, {\rm CPL}}=-0.54^{+0.83}_{-0.70}$ and peak energy of $E_{\rm peak}=78^{+87}_{-18}$~keV (90\% confidence) in the $15$--$150$~keV energy range. Adopting the CPL model parameters and a redshift of $z=0.5536$ (Section~\ref{sec:hostobs}), we calculate an isotropic-equivalent $\gamma$-ray energy ($E_{\gamma,{\rm iso}}$) of $E_{\gamma,{\rm iso}}$(15-150~keV)=$(8.4 \pm 1.1) \times 10^{49}$~erg.

\subsection{Swift X-ray observations}

We re-analyze the \swift\ XRT observations of \grb\ to obtain the X-ray light curve spanning $\delta t \approx 0.006-2.74$~days. To perform the X-ray spectral analysis, we obtain the source and background spectra, ancillary and response files for each bin of the light curve as defined by the XRT time-sliced spectra interface \citep{ebp+09}. We reduced the data using the \texttt{HEASoft} software \citep[v.6.26.1;][]{Blackburn1999,NASA2014} and \texttt{caldb} files (v.\,20190910). We use the methods of \citet{ebp+07} and \citet{ebp+09} for selecting the source and background regions and binning the data, as well as for extracting the counts and producing the spectra.

We first use the \texttt{Xspec} software \citep[v.12.9.0;][]{arn96} to fit the spectrum of each bin of the light curve (0.3-10~keV), binning the spectra using \texttt{grppha} to ensure a minimum of one count per bin. We use \texttt{VERN} X-ray cross-sections \citep{vfk+96}, \texttt{WILM} abundances \citep{wam00} and W-statistics for background-subtracted Poisson data \citep{wlk79}. We employ a two-component absorption power-law model characterized by photon index ($\Gamma_X$), the intrinsic hydrogen column density ($N_{\rm H, int}$) at the redshift of the GRB (see Section~\ref{sec:hostobs}), and the Galactic Hydrogen column density in the direction of the \grb ($N_{\rm H, MW}=2.94\times10^{20}$~cm$^{-2}$; \citealt{wsb+13}). Allowing both $\Gamma_X$ and $N_{\rm H, int}$ to vary, we find that the value of $N_{\rm H, int}$ is consistent with zero, and that the individual values for $\Gamma_X$ do not exhibit statistically significant changes (to within $1\sigma$) over the course of the observations.

Since the parameter values for the individual observations are poorly constrained, we use {\tt Xspec} to jointly fit the entire data set, and find best-fit values of $\Gamma_X=1.47^{+0.24}_{-0.19}$ ($1\sigma$ confidence) and $N_{\rm H, int}<5.51 \times 10^{21}$~cm$^{-2}$ ($3\sigma$). Fixing the spectral parameters to the best-fit values and freezing $N_{\rm H, int}=0$~cm$^{2}$, we calculate the unabsorbed X-ray fluxes utilizing the {\tt cflux} model within the 0.3-10~keV energy range. Finally, we determine the X-ray afterglow flux densities, $F_{\nu,X}$ at $\nu_X=1$~keV, using the spectral index, $\beta_X$ ($\beta_X\equiv 1-\Gamma_X$) which has a value of $\beta_X =-0.47^{+0.24}_{-0.19}$ across all observations.

For the last observation at $\delta t = 2.74$~days, we determine the $3\sigma$ count-rate upper limit using the four source photons detected in $\sim 4.8$~ks using Poissonian statistics following \citet{geh86}. Applying the best-fit spectral parameters using {\tt WebPIMMS}\footnote{\url{https://heasarc.gsfc.nasa.gov/cgi-bin/Tools/w3pimms/w3pimms.pl}}, we calculate the unabsorbed X-ray flux and resulting upper limit on $F_{\nu,X}$. The observational details, 1~keV flux densities and $1\sigma$ uncertainties for the entire X-ray afterglow light curve are listed in Table~\ref{tab:obs}. These results are consistent within $1\sigma$ uncertainties to the {\it Swift} time-sliced interface results \citep{ebp+09} under the same assumptions in spectral binning.

\begin{deluxetable*}{ccccccccccccc}[t!]
\tablecaption{Broad-band Afterglow and Host Galaxy Observations of GRB\,200522A \label{tab:obs}}
\tablecolumns{10}
\tablewidth{0pt}
\tablehead{
\colhead{$\delta t^{a}$} &
\colhead{Band} & 
\colhead{Facility} & 
\colhead{Instrument} &
\colhead{Exp. time} & 
\colhead{Afterglow} & 
\colhead{Afterglow} & 
\colhead{Host Galaxy} & 
\colhead{A$_{\lambda}$} &
\colhead{Ref.} & \\
\colhead{(days)} & 
\colhead{} &
\colhead{} & 
\colhead{} & 
\colhead{(s)} & 
\colhead{(AB mag)} & 
\colhead{($\mu$Jy)} & 
\colhead{(AB mag)} & 
\colhead{(AB mag)} &
\colhead{}
}
\startdata
\hline
\multicolumn{10}{c}{\emph{X-rays}} \\
\hline
0.0059 & 1~keV & \swift & XRT & 232.2 & \nod &	$0.34 \pm 0.080$ & \nod & \nod & 1 \\
0.048  & 1~keV & \swift & XRT &	492.0 & \nod &	$0.14 \pm 0.036$ & \nod & \nod  & 1\\
0.056  & 1~keV & \swift & XRT &	871.6 & \nod &	$0.15 \pm 0.028$ & \nod & \nod & 1 \\
0.16   & 1~keV & \swift & XRT &	2105.0 & \nod &	$0.036 \pm 0.0091$ & \nod & \nod & 1 \\
0.64   & 1~keV & \swift & XRT &	8890.0	& \nod & $0.017 \pm	0.0031$ & \nod & \nod  & 1\\
2.74   & 1~keV & \swift & XRT &	4834.1 & \nod & $<0.011$ & \nod & \nod & 1 \\
\hline
\multicolumn{10}{c}{\emph{Optical/NIR}} \\
\hline
0.28 & clear & BOOTES-3 &  & 900 & $>18.1$ & $<208.9$ & \nod & 0.066 & 2 \\
0.65 & $R$ & LCOGT & Sinistro & 900 & $\gtrsim 22.1$ & $\lesssim 5.25$ & $21.27 \pm 0.17$ & 0.059 & 1, 3 \\
0.69 & $I$ & LCOGT & Sinistro & 900 & $>20.4$ & $<25.35$ & \nod & 0.041 & 1, 3 \\
2.12 & $r$ & Gemini-N & GMOS & 630 & $>22.2^b$ & $<4.78$ & $21.31 \pm 0.10$ & 0.062 & 4 \\
3.52 & F125W & HST & WFC3 & 5223.5 & $24.53 \pm 0.15$ & $0.55 \pm 0.07$ & $20.95 \pm 0.01$ & 0.020 & 1 \\
3.66 & F160W & HST & WFC3 & 5223.5 & $24.61 \pm 0.15$ & $0.51 \pm 0.07$ & $20.65 \pm 0.01$ & 0.014 & 1 \\
16.38 & F125W & HST & WFC3 & 4823.5 & $>27.5$ & $<0.036$ & $20.84 \pm 0.01$ & 0.020 & 1 \\
30.09 & $G$ & Keck & LRIS & 480 & \nod & \nod & $22.18 \pm 0.02$ & 0.090 & 1 \\
30.09 & $R$ & Keck & LRIS & 360 & \nod & \nod & $21.14 \pm 0.02$ & 0.059 & 1 \\
32.60 & $R$ & LCOGT & Sinistro & 1200 & \nod & \nod & $21.97 \pm 0.18$ & 0.059 & 1 \\
55.24 & F125W & HST & WFC3 & 5223.5 & \nod & \nod & $20.84 \pm 0.01$ & 0.020 & 1 \\
55.37 & F160W & HST & WFC3 & 5223.5 & \nod & \nod & $20.67 \pm 0.01$ & 0.014 & 1 \\
56.12 & $Z$ & Keck & DEIMOS & 960 & \nod & \nod & $20.84 \pm 0.01$ & 0.034 & 1 \\
56.13 & $I$ & Keck & DEIMOS & 960 & \nod & \nod & $20.93 \pm 0.01$ & 0.041 & 1 \\
56.14 & $V$ & Keck & DEIMOS & 480 & \nod & \nod & $>21.26$ & 0.075 & 1 \\
Archival & $u$ & SDSS & & \nod & \nod & \nod & $22.43 \pm 0.31$ & 0.116 & 5 \\
Archival$^{c}$ & $y$ & PS1 & & \nod & \nod & \nod & $20.87 \pm 0.30$ & 0.030 & 1, 6 \\
Archival$^{c}$ & 3.6 $\mu$m & Spitzer & \nod & \nod & \nod & \nod & $21.07 \pm 0.10$ & \nod & 1, 7-8 \\
Archival$^{c}$ & 4.5 $\mu$m & Spitzer & \nod & \nod & \nod & \nod & $21.30 \pm 0.10$ & \nod & 1, 7-8 \\
\hline
\multicolumn{10}{c}{\emph{Radio}} \\
\hline
0.23 & 6.05~GHz & VLA & & 2700 & \nod & $33.4 \pm 8.2$ & \nod & \nod & 1  \\
2.19 & 6.05~GHz & VLA & & 2640 & \nod & $27.1 \pm 7.2$ & \nod & \nod & 1  \\
2.19 & 9.77~GHz & VLA & & 2220 & \nod & $\lesssim 23.7$ & \nod & \nod & 1  \\
6.15 & 6.05~GHz & VLA & & 3720 & \nod & $\lesssim 18.6$ & \nod & \nod & 1  \\
11.15 & 6.05~GHz & VLA & & 5340 & \nod & $\lesssim 14.1$ & \nod & \nod & 1  \\ \\
1.21$^{d}$ & 6.05~GHz & VLA & & 5340 & \nod & $29.7 \pm 5.4$ & \nod & \nod & 1  \\
8.65$^{e}$ & 6.05~GHz & VLA & & 9060 & \nod & $\lesssim 10.9$ & \nod & \nod & 1  \\
\hline
\enddata
\tablecomments{All magnitudes are in the AB system and corrected for Galactic extinction in the direction of the burst, $A_{\lambda}$ \citep{sf11}. Uncertainties correspond to $1\sigma$ confidence and upper limits correspond to $3\sigma$. \\
$^a$ Mid-time of observation in the observer frame. \\
$^{b}$ Reported image limit within the XRT error region, outside of the host galaxy. \\
$^{c}$ These photometric points are a result of forced photometry at the position of the host galaxy in archival imaging. The host galaxy is uncatalogued in these bands. \\
$^{d}$ Combination of 6.05~GHz observations at 0.23~days and 2.19~days. \\
$^{e}$ Combination of 6.05~GHz observations at 2.19~days and 6.15~days. \\
{\bf References:} (1) This work; (2) \citealt{gcn27784}; (3) \citealt{gcn27794}; (4) \citealt{gcn27822}; (5) \citealt{aaa+15}; (6) \citealt{cmm+16}; (7) \citealt{Papovich2016}; (8) \citealt{Timlin2016} }
\end{deluxetable*}

\subsection{Optical follow-up observations}
\label{sec:opt}

The UltraViolet-Optical Telescope (UVOT) on-board \swift\ began observations of \grb\ at $\delta t = 448$~s, and obtained preliminary $3\sigma$ upper limits of $>19.5$~mag in the $white$ filter \citep{gcn27783}. Additional observations were taken with the Yock-Allen BOOTES-3 telescope starting at $\delta t \approx 6.8$~hr \citep{gcn27784} with an upper limit of $>18.1$~mag in the clear filter.

We initiated $R$ and $I$-band observations with the Sinistro instrument mounted on the Las Cumbres Observatory Global Telescope network (LCOGT) 1-meter telescope at the South African Astronomical Observatory for a total of 900~s of exposure time in each filter at mid-times of $\delta t = 0.65$ and $0.69$~days, respectively. These observations were first reported in \citet{gcn27792,gcn27794}, and the following analyses supercede those reported in the circulars. We reduce the data with the BANZAI\footnote{\url{https://github.com/LCOGT/banzai}} data reduction pipeline, which performs bad-pixel masking, bias subtraction, dark subtraction, flat field correction, source extraction (using SEP, the Python and C library for Source Extraction and Photometry), and astrometric calibration (using astrometry.net). We align the frames and co-add the individual images using Python/{\tt astroalign}, and perform astrometry relative to the USNO-B1 catalog.

Within the XRT position, we detect a single, clear source in the images, consistent with the position of the SDSS catalogued galaxy SDSSJ002243.71-001657.5 \citep{aaa+15}, first reported as the potential host galaxy in \citet{gcn27779}. Performing photometry with {\tt SExtractor} relative to USNO-B1.0, we calculate a magnitude of $R=21.27 \pm 0.17$~mag, consistent with the archival SDSS magnitude of $r=21.17 \pm 0.07$~mag, and an upper limit of $I\gtrsim 20.39$~mag within the XRT position (Table~\ref{tab:obs}). 

We obtained a second, deeper set of LCO $R$-band observations at $\delta t \approx 32.6$~days. Performing image subtraction between the two LCO epochs using the {\tt HOTPANTS} software package \citep{bec15}, we do not find any significant residuals. We thus measure a $3\sigma$ upper limit on optical afterglow emission of $R\gtrsim 22.1$~mag at $\delta t \approx 0.65$~days. The details of our observations are listed in Table~\ref{tab:obs}. We note that reported observations taken with the Gemini Multi-Object Spectrograph (GMOS) mounted on the Gemini-North telescope also place a comparable limit on emission outside of the host galaxy but within the XRT position of $r>22.2$~mag \citep{gcn27822}. 

\subsection{Radio afterglow discovery}
\label{sec:radio}

\begin{figure*}
\centering
\includegraphics[width=0.7\textwidth,trim={0 1in 0 0}]{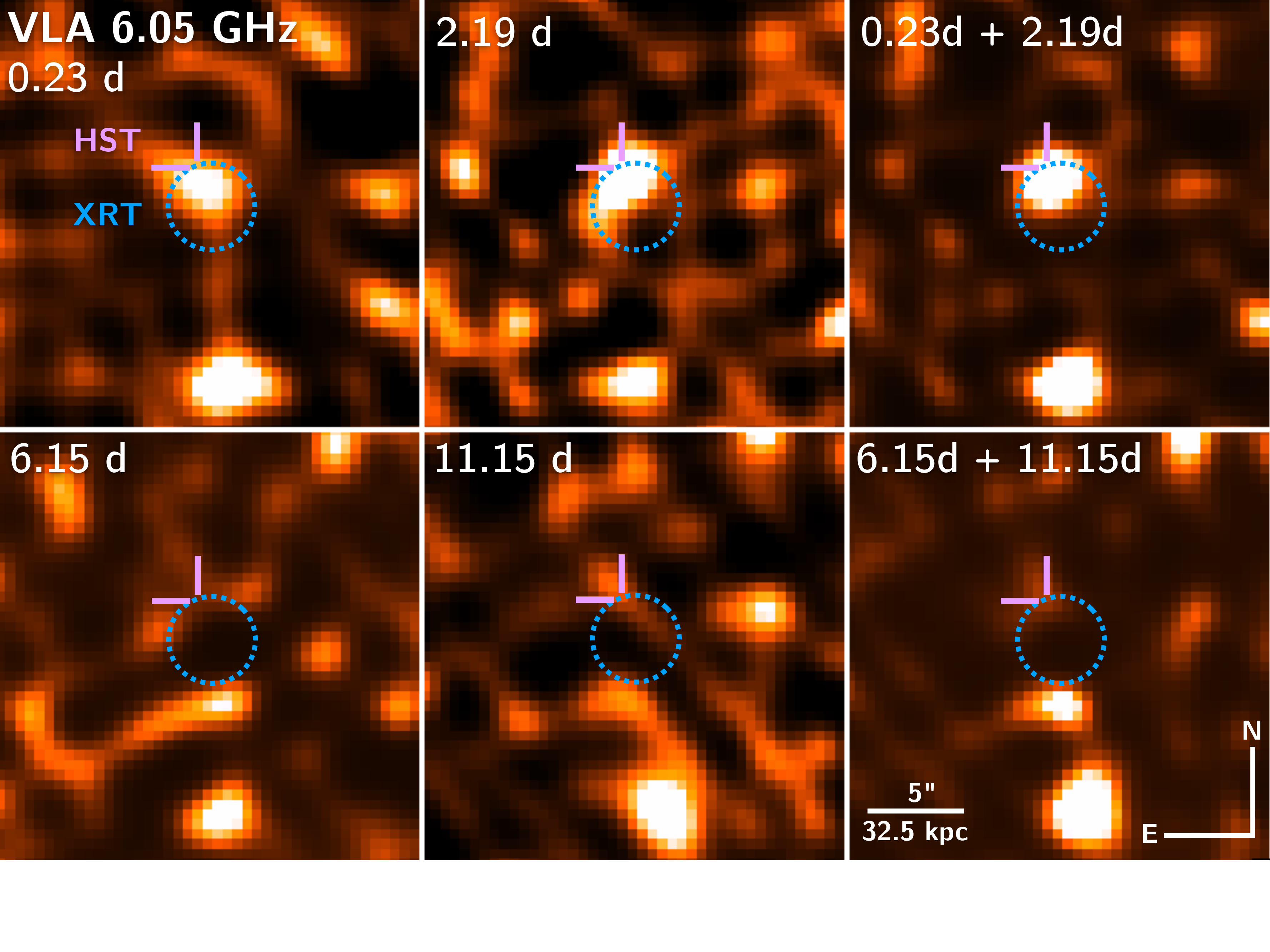}
\caption{VLA observations revealing the radio afterglow of \grb. The first two columns represent the four epochs of VLA observations at 6.05~GHz (C-band) taken at $\delta t = 0.23$, $2.19$, $6.15$, and $11.15$~days, respectively. The final column represents combined observations of the first two epochs and the final two epochs, in which a fading source within the XRT position (blue dotted; 90\% confidence) is apparent. The {\it HST} NIR counterpart position (purple cross-hairs) is also denoted in each panel. The scale and orientation of all panels is displayed in the bottom-right panel.
\label{fig:vla_ag}}
\end{figure*}

We initiated observations with the Karl G. Jansky Very Large Array (VLA; Program 19B-217; PI:~Fong; reported in \citealt{gcn27786}) at a central frequency of 6.05~GHz (C-band). The observations occurred at a mid-time of $\delta t = 0.23$~days for a total of 1~hr, including time for flux density and phase calibration. We centered the upper and lower sidebands at 5.0~GHz and 7.2~GHz, respectively, and used 3C147 for flux calibration and J0022+0014 for gain calibration. We excised the effects of radio frequency interference (RFI) from the data, and employed standard interferometric calibration techniques for data calibration and analysis within the Common Astronomy Software Applications (CASA; \citealt{CASA}). We used CASA/{\tt tclean} to image the field, employing Briggs weighting with a robust parameter of 0 (to minimize side-lobe contamination from neighboring sources) and two Taylor terms (nterms$=2$). Toward the Northeast edge of the 90\% XRT error circle, we detect a single radio source (Figure~\ref{fig:vla_ag}). Using a point source model within CASA/{\tt imfit}, we measure a source flux density of $F_{\nu,{\rm 6 GHz}}=33.4 \pm 8.2\,\mu$Jy.

We obtained a second 6.05~GHz epoch at $\delta t=2.19$~days, in which the source is still detected with $F_{\nu,{\rm 6 GHz}}=27.1 \pm 7.2\,\mu$Jy, consistent with a constant flux density within the $1\sigma$ errors. In addition, we obtained contemporaneous observations at a mean frequency of $9.77$~GHz, and do not detect any significant emission within the X-ray error circle to a $3\sigma$ limit of $F_{\nu,{\rm 9.7 GHz}}\lesssim 23.7\,\mu$Jy. To assess the nature of the source at $6.05$~GHz within the XRT error circle, we obtained a final series of deeper observations at $6.05$~GHz at $\delta t \approx 6.15$ and $11.15$~days. The source is no longer detected to $3\sigma$ limits of $F_{\nu,{\rm 6~GHz}}\lesssim 18.6\,\mu$Jy and $\lesssim 14.1\,\mu$Jy, respectively.

We use CASA/{\tt concat} to combine the exposures of the first two C-band epochs, and derive a position of RA=\ra{00}{22}{43.706}, Dec=$-$\dec{00}{16}{57.97} (J2000) with $1\sigma$ positional uncertainties of $\Delta$RA$=0.23''$ and $\Delta$Dec=$0.27''$, with a flux density of $F_{\nu,{\rm 6 GHz}}=29.7 \pm 5.3\,\mu$Jy. Combining the final two observations in the same manner, we determine a deep limit of $F_{\nu,{\rm 6 GHz}}\lesssim 10.9\,\mu$Jy ($3\sigma$). Due to the spatial coincidence with the XRT and {\it HST} NIR counterpart positions (see Section~\ref{sec:hst}), along with clear fading behavior of the source, we consider this to be the radio afterglow of \grb. The individual epochs and combined images are displayed in Figure~\ref{fig:vla_ag} and the details of our observations are summarized in Table~\ref{tab:obs}.

\subsection{Hubble Space Telescope NIR counterpart discovery}
\label{sec:hst}

\begin{figure*}
\centering
\includegraphics[width=\textwidth,trim={0 2.8in 0 0}]{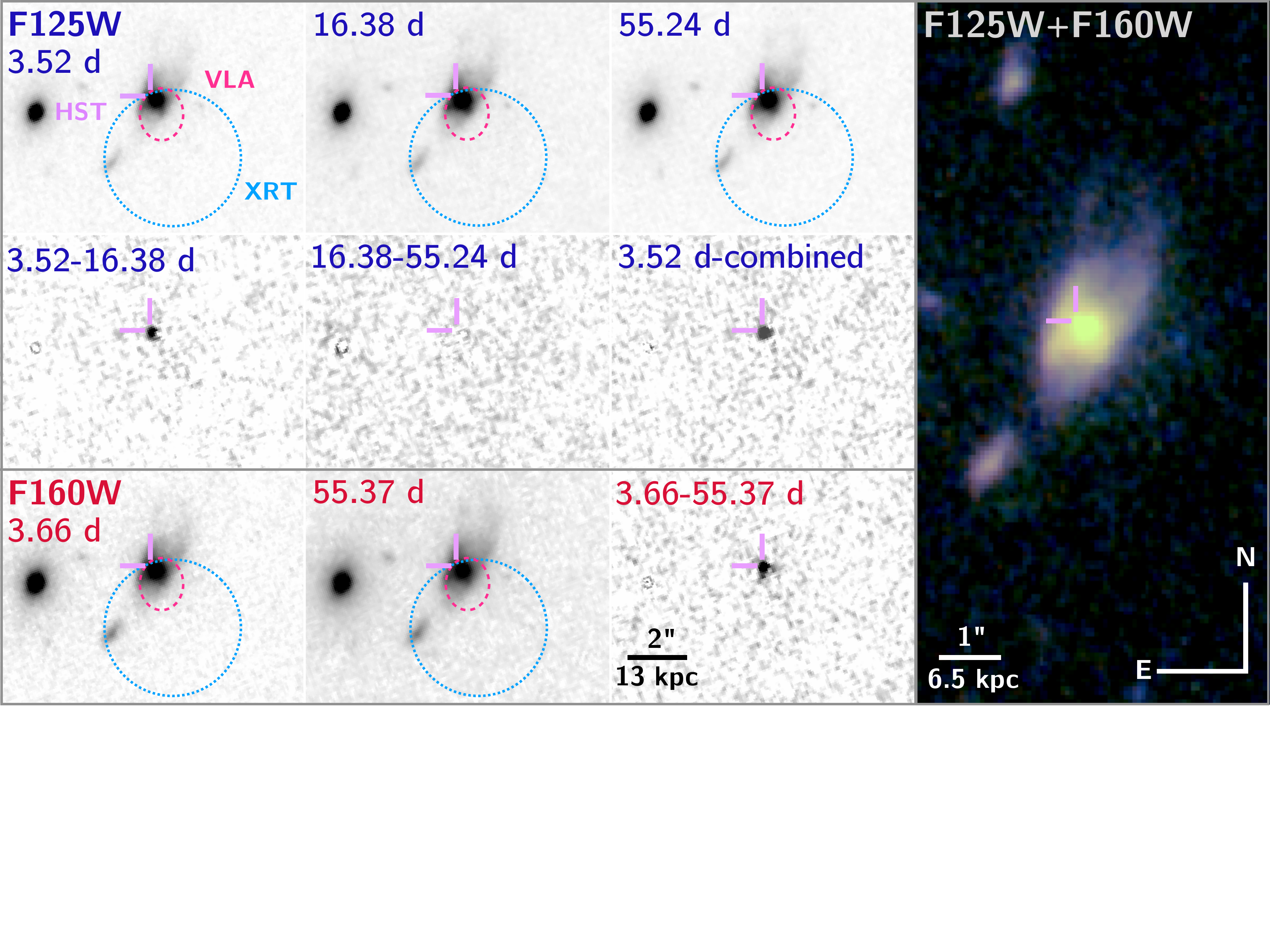}
\caption{{\it HST}/WFC3 observations of \grb. The three epochs of F125W observations are displayed (top), along with the corresponding {\tt HOTPANTS} residual images (middle); ``combined'' refers to a merged template of the F125W observations at 16.38 and 55.24 days. The residual images reveal a fading source between 3.52 and 16.38 days. The two epochs of F160W observations and the subtraction between the two visits are shown in the bottom row. In each of the smaller panels, the XRT position (blue dotted; $90\%$ confidence), VLA position (pink dashed ellipse; $3\sigma$), and {\it HST} NIR counterpart position (purple cross-hairs) are shown. The scale is denoted in the bottom right panel. The right-hand image is a color composite composed of the merged F125W template and F160W images, with the position of the {\it HST} counterpart denoted by the purple cross-hairs.
\label{fig:optpanel}}
\end{figure*}

We initiated observations with the {\it Hubble Space Telescope} ({\it HST}; PI: Berger, Program 15964) using the Wide Field Camera~3 (WFC3) IR channel (previously reported in \citealt{gcn27826,gcn27827}). We obtained observations in the F125W and F160W bands for a total of 5223.5~s in each filter at mid-times of $\delta t = 3.52$~days and $3.66$~days, respectively. We used the {\tt astrodrizzle} package to combine the images in each filter, employing {\tt combine\_type}=median, {\tt wht\_type}=EXP, {\tt pixscale}$=0.0642''$~pixel$^{-1}$ (half of the native WFC3/IR pixel scale) and {\tt pixfrac}=0.8. The images are shown in Figure~\ref{fig:optpanel}. We performed absolute astrometry on the F125W filter image relative to SDSS DR12 \citep{aaa+15}, with an astrometric tie uncertainty of $0.048''$ ($1\sigma$). The host galaxy (Section~\ref{sec:hostobs}) is clearly detected at a position of RA=\ra{00}{22}{43.717}, Dec=$-$\dec{00}{16}{57.46}, along with an additional fainter, extended source within the XRT error circle to the southeast at RA=\ra{00}{22}{43.813}, Dec=$-$\dec{00}{16}{59.52}. We also note the presence of a point source $\approx 1.43''$ to the east of the host galaxy (Figure~\ref{fig:optpanel}; \citealt{gcn27827}).

We obtained two additional sets of observations in the F125W filter at mid-times of $\delta t = 16.38$~days and $55.24$~days \citep{gcn27904}, and one additional set in the F160W filter at $55.37$~days, which we treat in the same manner as the first epoch. For each observation, we used IRAF/{\tt ccmap} and {\tt ccsetwcs} to perform astrometry relative to the first epoch of F125W observations (which itself is tied to SDSS), with an average relative astrometric uncertainty of $\approx 0.01''$.

Using the observations at $\delta t \approx 55$~days as a template for each filter, we performed image subtraction using the {\tt HOTPANTS} software package \citep{bec15} between each of the earlier epochs and the template in the relevant filter. The difference images at $\delta t \approx 3.6$~days reveal a point source present at the Northeast edge of the XRT position,  consistent with the radio afterglow position with RA = \ra{00}{22}{43.727}, Dec=$-$\dec{00}{16}{57.43} (Figure~\ref{fig:optpanel}) in both filters. This source subsequently fades in F125W imaging by $16.4$~days. Given the fading behavior and coincidence with the X-ray and radio positions, we consider this source to be the NIR counterpart to \grb.

The lack of residuals in the difference image between the latter two F125W epochs signifies a negligible amount of transient emission at $\delta t=16.38$~days. Thus, we use {\tt astrodrizzle} to create a ``combined'', deep F125W template. The results of the image subtraction between the first epoch and the deep template are shown in Figure~\ref{fig:optpanel}, exhibiting a high-significance detection of the NIR counterpart, on which we base our subsequent photometry.

The difference images all exhibit contamination coincident with the core of the host galaxy. Each sub-frame in the first set of observations have ${\tt EXPTIME}=602.93$~s with peak counts near the center of the galaxy of $\approx$4200~e$^{-}$. This means that the center of the galaxy is non-linear at the $0.1\%$ level\footnote{\url{http://documents.stsci.edu/hst/wfc3/documents/handbooks/}}, and that even with non-linearity corrections, this will result in an imperfect subtraction at the host centroid.

To obtain reliable photometry and uncertainties of the NIR counterpart, we pursue three independent methods: (1) aperture photometry using a small aperture with an encircled energy (EE) correction, (2) PSF photometry with width fixed to the in-band WFC3/IR PSF, and (3) PSF photometry with an empirically-determined value. First, using the IRAF/{\tt phot} package, we perform aperture photometry of the source using a small, $0.2''$-radius aperture fixed at the position of the counterpart. We then apply tabulated encircled energy corrections to correct the small apertures to infinity\footnote{https://www.stsci.edu/hst/instrumentation/wfc3/data-analysis/photometric-calibration/ir-encircled-energy}, with corrections of $0.29$ (F125W) and $0.34$~mag (F160W). For the second method, we use the tabulated values of the FWHM WFC3/IR PSF \citep{Windhorst11} of 0.136\arcsec\ for the F125W filter and 0.150\arcsec\ for F160W.  We then construct a fixed-width Gaussian PSF using {\tt photutils} and apply it in a 0.5\arcsec\ aperture at the location of the residual in our F125W and F160W difference images, fitting for the integrated flux and centroid position of the source in both images.  We derive our uncertainties on flux by changing the best-fitting centroid and fixed-width FWHM to within 10\% of the input values and measuring the standard deviation in the implied flux. For the third method, we use {\tt daophot} to empirically determine the best-fit PSF size and shape from isolated stars in the epoch one images.  With the resulting PSF model, we then fit for the integrated flux and centroid of the residual in the difference images. Taking the average flux and statistical uncertainty of the results from the three methods in flux-space, we find that the NIR counterpart brightness is $m_{\rm F125W}=24.55 \pm 0.15$~mag and $m_{\rm F160W}=24.62 \pm 0.15$~mag, in which the dominant source of uncertainty is the difference in methods (with individual measurement uncertainties of $\lesssim 0.05$~mag).

Finally, to obtain an upper limit in the $16.38$~day observation, we use {\tt dolphot} to inject fake sources of known brightness ($m_{\rm F125W}=24-28.5$~mag) at and near the counterpart location in the difference image. These sources have a shape matched to the WFC3/IR F125W instrumental PSF. We then recover these sources using {\tt dolphot} and change the brightness in increments of 0.1~mag until we find the threshold at which $>$99.7\% of sources are recovered at a signal-to-noise of $>$3, from which we derive $m_{\rm F125W}\gtrsim27.5$~mag ($3\sigma$) at $\delta t \approx 16.38$~days.

\subsection{Host galaxy observations and redshift}
\label{sec:hostobs}

\begin{figure*}
\centering
\includegraphics[width=0.6\textwidth]{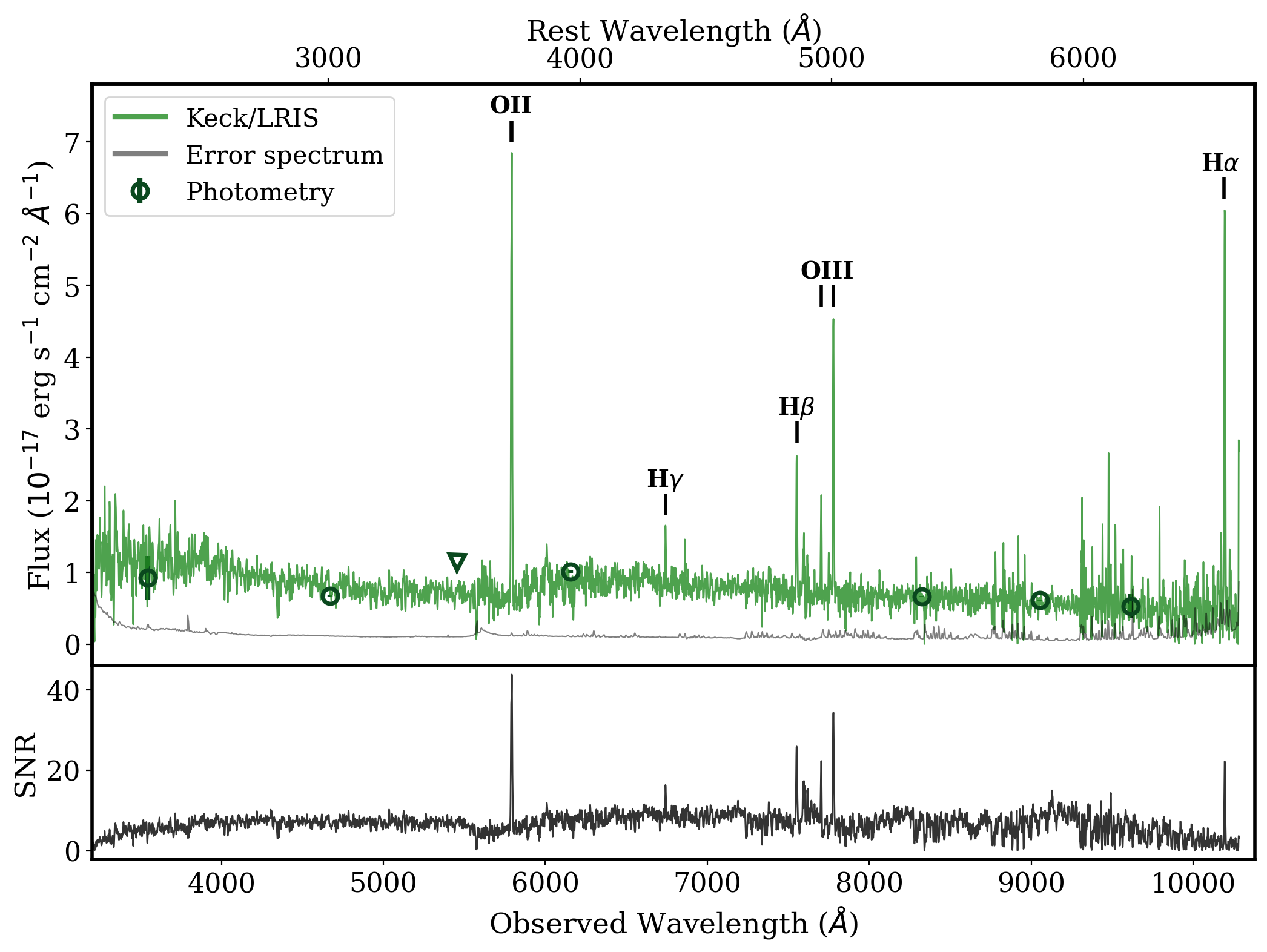}
\vspace{-0.1in}
\caption{{\it Top:} Keck/LRIS spectrum of the host galaxy of \grb\ (green) and error spectrum (grey) along with the $uGVRIZy$-band photometry (dark green circles for detections and triangle for upper limit) from SDSS DR12 \citep{aaa+15}, Pan-STARRS \citep{cmm+16}, Keck/LRIS and Keck/DEIMOS. The location of prominent emission lines are marked. {\it Bottom:} Signal-to-noise ratio (SNR) of the spectrum, showing the overall continuum shape, and the high SNR of marked emission lines. 
\label{fig:spec}}
\end{figure*}

To quantify the probability that SDSSJ002243.71-001657.5 is the host galaxy of \grb, we calculate the angular offsets between the NIR counterpart and the host galaxy centroid derived in {\it HST} imaging. We use the final observations at $\delta t \approx 55$~days, as the host centroid determination in earlier epochs will be contaminated by the transient emission. We consider three sources of uncertainty in the offset calculation: the counterpart positional uncertainty ($\sigma_{\rm HST}=0.0012''$), the host positional uncertainty ($\sigma_{\rm host, F125W}=0.052''$, $\sigma_{\rm host, F160W}=0.0007''$) and the relative astrometric uncertainties between {\it HST} observations ($\sigma_{\rm tie, F125W}=0.029''$, $\sigma_{\rm tie, F160W}=0.013''$). We measure projected angular offsets of $\delta R = 0.155 \pm 0.054''$ (F125W) and $0.143 \pm 0.029''$ (F160W). Using the angular offsets and $R$-band magnitude of the host galaxy (Table~\ref{tab:obs}), we calculate a low probability of chance coincidence of $P_{\rm cc}=3.5 \times 10^{-5}$ following the methods of \citet{bkd02}. There are only two other catalogued galaxies within $0.5'$, both of which have significantly higher values of $P_{\rm cc}=0.25$--$0.4$. Repeating the same exercise based on the VLA position, and taking into account the absolute astrometric uncertainty between the F125W observations and SDSS DR12, we calculate a similarly low $P_{cc} = 4.8 \times 10^{-4}$. We thus confirm SDSSJ002243.71-001657.5 as the host galaxy of \grb.

To further characterize the host galaxy, we used the Low Resolution Imaging Spectrometer (LRIS) mounted on the 10-m Keck I telescope (PI:~Blanchard; Program O287) to obtain $G$- and $R$-band imaging on 2020 Jun 21 UT at a mid-time of $\delta t \approx 30.1$~days (Table~\ref{tab:obs}). We apply bias and flat-field corrections using the {\tt photpipe} image reduction and processing software \citep{Rest05,Kilpatrick18:16cfr}.  We perform relative alignment of the individual frames and stack them with the {\tt SWarp} software package \citep{SWarp}. For the final stacked frames, we use IRAF tasks {\tt ccmap} and {\tt ccsetwcs} to align the images to SDSS DR12.

We also obtained $I$-, $Z$- and $V$-band imaging of the host galaxy with the DEep Imaging Multi-Object Spectrograph (DEIMOS) mounted on the 10-m Keck II telescope on 2020 Jul 17 UT at a mid-time of $\delta t \approx 56.1$~days (Table~\ref{tab:obs}; PI:~Blanchard). We apply bias and flat-field corrections, and align and stack the individual images using a custom pipeline\footnote{\url{https://github.com/CIERA-Transients/Imaging_pipelines/blob/master/DEIMOS_pipeline.py}}. We perform aperture photometry using {\tt phot}, employing source apertures of $2.5''$, chosen to fully encompass the host galaxy. After calibrating each image to the SDSS DR12 catalog and converting to the AB system using the relevant relations from \cite{Chonis2008}, we obtain host galaxy magnitudes in the $GRIZ$ filters, and an upper limit in the $V$-filter; the results are listed in Table~\ref{tab:obs}. From our {\it HST} imaging (Section~\ref{sec:hst}), we use IRAF/{\tt phot} to measure host magnitudes of $m_{\rm F125W}=20.84 \pm 0.01$~mag and $m_{\rm F160W}=20.65 \pm 0.01$~mag (Table~\ref{tab:obs}). 

We supplement these data with available photometry in other bands based on archival imaging in the SDSS DR12, Pan-STARRS1 (PS1), and {\it Spitzer} Space Telescope imaging as part of the Stripe 82 survey (Program 90053, PI: Richards; \citealt{aaa+15,cmm+16,Spitzer2004,Timlin2016,Papovich2016}). For SDSS DR12, the host galaxy is catalogued and we use the available $u$-band photometry to supplement the Keck photometry. The host galaxy is weakly detected in the PS1 3$\pi$ $y$-band stacks, and in the {\it Spitzer} $3.5\,\mu$m and $4.6\,\mu$m imaging, but is not catalogued. Thus, we download the imaging and perform aperture photometry of the host. The {\it Spitzer} photometry is complicated by a varying background due to nearby sources, which we ameliorate by selecting $\approx5$ source-free, background regions in the vicinity of the host, and report the variance in the derived flux density as the uncertainty. Our host galaxy photometry based on archival imaging is also listed in Table~\ref{tab:obs}.

In addition, we obtained Keck/LRIS spectroscopy on 2020 Jun 21 UT for a total of $3\times 900$~s with the blue camera, and $3\times860$~s with the red camera, with a fixed dichroic wavelength of $5600$\AA. The spectrum was taken with a $1.0''$ longslit, 400/3400 grism (blue) and the 400/8500 grating (red), with a central wavelength of 7830~\AA. The resulting spectrum spans a continuous range of $\approx 3200-10280$~\AA\ with a spectral resolution of $\sim7$~\AA\ in both arms. We use standard {\tt IRAF} tasks to subtract the overscan, apply flat-field corrections, model the sky background and subtract it for the individual frames. We also perform wavelength calibration using HeNeArCdZn arc lamp spectra, and spectrophotometric flux calibration using the standard star Feige110 taken at a similar airmass on the same night. We use {\tt apall} to extract the 1D spectra, which we then co-add. We determine the error spectrum by performing the same reduction steps but on spectra without sky subtraction and performing standard error propagation in the combination. The resulting spectrum is shown in Figure~\ref{fig:spec}.

The spectrum overall exhibits a blue continuum, with a 4000\AA\ break at $\sim$5800\AA. We detect several emission lines: [OII]$\lambda3727$, [OIII]$\lambda4959,5007$, and the Balmer lines H$\alpha$, H$\beta$, and H$\gamma$. Cross-correlating the host spectrum of \grb\ to a star-forming galaxy template as part of the SDSS~DR5 template library \citep{SDSS-DR5}, we calculate a common redshift and $1\sigma$ uncertainty of $z=0.5536 \pm 0.0003$. At this redshift, the projected physical offset of the NIR counterpart to \grb\ in the F160W filter is $\delta R = 0.93 \pm 0.19$~kpc.

\section{Broad-band Modeling I: A Forward Shock with a NIR Excess}
\label{sec:bb_nirexcess}

In the following two sections (Sections~\ref{sec:bb_nirexcess} and \ref{sec:bb_alternative}), we present our afterglow modeling and two interpretations of the broad-band data set (termed Scenarios I and II, respectively).

\begin{figure*}
\centering
 \begin{minipage}{\textwidth}
   \centering
   $\vcenter{\hbox{\includegraphics[width=0.49\textwidth]{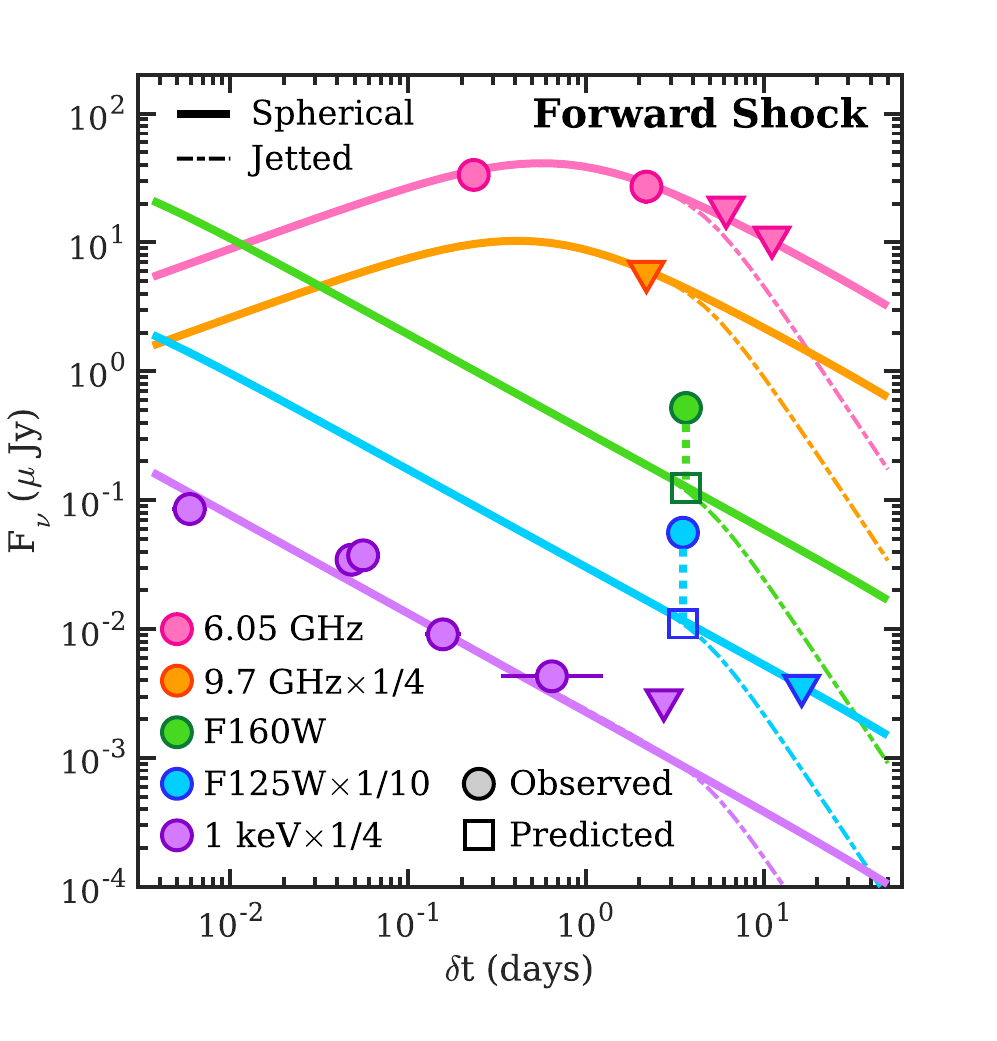}}}$
   $\vcenter{\hbox{\includegraphics[width=0.49\textwidth]{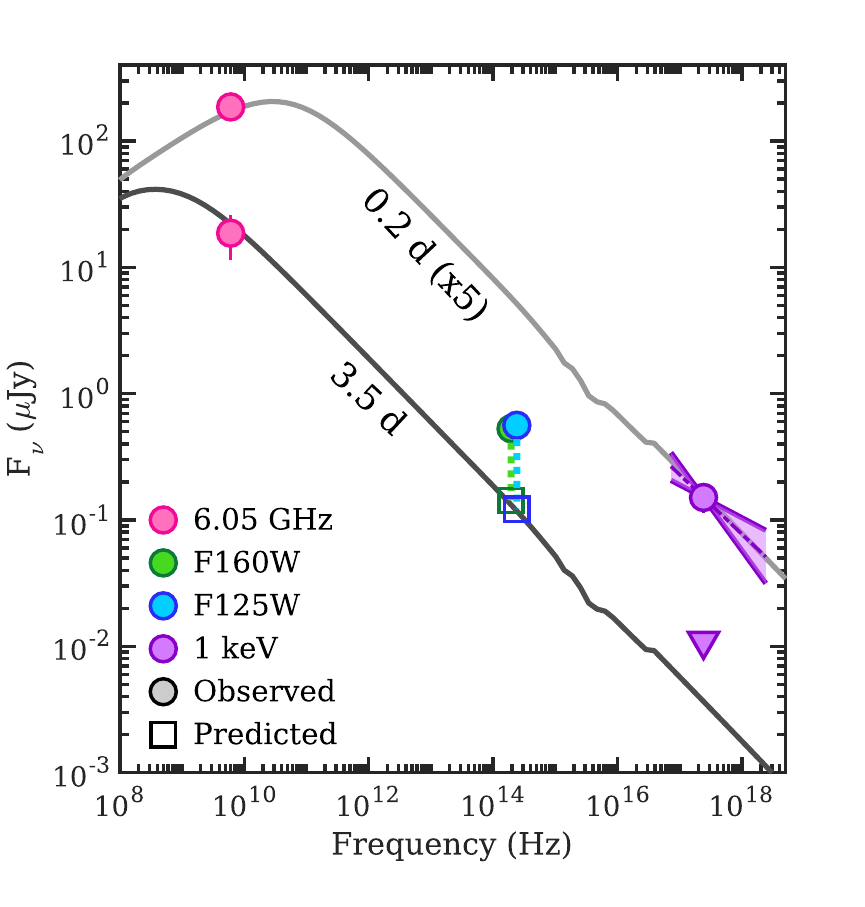}}}$
 \end{minipage}
\vspace{-0.2in}
\caption{The radio, NIR and X-ray observations of the counterpart of \grb\ (circular points) and models in Scenario I. {\it Left:} Representative afterglow model light curves representing a forward shock propagating into the circumburst medium for a spherical outflow (solid lines) and a jetted outflow (dot-dashed lines). If a jet break exists, the observations constrain the time of the break to $\delta t \gtrsim 3.5$~days. {\it Right:} The corresponding afterglow model's spectral energy distributions at $\delta t = 0.2$~days and $3.5$~days; jetted and spherical models are the same at these times. In both panels, models and data points are scaled as denoted for clarity. Error bars correspond to $1\sigma$ and are generally smaller than the size of the symbols, and triangles correspond to $3\sigma$ upper limits. The radio and X-ray afterglow temporal and spectral evolution are consistent with the forward shock model, and the measured X-ray spectral slope (purple regions, representing $1\sigma$ confidence region) is in agreement with the model. Meanwhile, the observed F125W and F160W fluxes at $\delta t = 3.52$ and $3.66$~days are in excess of the predicted fluxes (open squares) by factors of $\approx 5-10$.
\label{fig:aglc}}
\end{figure*}

\subsection{Model Description}

Here, we first interpret the radio, NIR, and X-ray observations of \grb\ in the context of synchrotron emission from a forward shock (FS) produced by the interaction of the GRB jet with the ambient environment \citep{spn98,gs02}. The parameters of the model are the isotropic-equivalent kinetic energy (\EKiso) of the jet, the particle density of the circumburst environment (\dens), the power-law index of accelerated electrons ($p$), the opening angle of the outflow (\thetajet), and the fractions of the forward shock energy imparted to electrons (\epse) and magnetic fields (\epsb). The resulting synchrotron spectrum is characterized by three break frequencies: the synchrotron self-absorption frequency (\nua), the characteristic synchrotron frequency (\numax), and the cooling frequency (\nuc). We use the convention $F_{\nu} \propto t^\alpha \nu^\beta$ throughout. 

We assume negligible intrinsic extinction, which is supported by the observed Balmer decrement in the Keck spectrum as consistent with the theoretical value, and the broad-band SED modeling of the host galaxy (Section \ref{sec:hostsps}). We also assume a uniform-density profile characteristic of the interstellar medium (ISM), as expected for short GRBs. 

At high electron Lorentz factors,  inverse-Compton (IC) cooling (with a strength determined by the Compton-$Y$ parameter) modifies the electron distribution and the resulting synchrotron radiation. Whereas IC cooling can be significant for long-duration GRBs \citep{se01,lbm+15}, for the typical parameters of short GRBs ($\EKiso\approx10^{51}$~erg, $\dens\approx10^{-2}~{\rm cm}^{-3}$; \citealt{fbm+15}), the Klein-Nishina (KN) effect limits $Y<Y_{\rm max}\approx0.2$ (assuming $p\approx2.2$ and $\epse\approx0.1$)\footnote{This limit, $Y_{\rm max}\propto t^{-\frac{5(p-2)}{2(p+2)}}$ is time-independent for $p\approx2$.}. In this regime, the synchrotron spectrum is better approximated by ignoring IC cooling effects \citep{nas09}. We therefore ignore IC cooling in our modeling, and subsequently verify whether the KN limit indeed applies to the derived parameters. 

From the XRT data, we measure $\beta_X=-0.47^{+0.24}_{-0.19}$ (Section~\ref{sec:obs}) and $\alpha_X=-0.67 \pm 0.10$ ($1\sigma$) over $\delta t\approx6\times10^{-3}$ days to 0.6~days. For the radio band, we measure a fairly shallow radio evolution of $\alpha_R=-0.1 \pm 0.2$ between $\delta t = 0.23$ and $2.19$~days, followed by a decline of $\alpha_R \lesssim -0.4$ at $\delta t > 2.2$~days. The faintness of the radio detection precludes a meaningful in-band spectral index. The non-detection at 9.77~GHz implies the radio emission is optically thin ($\beta_{\rm radio}\lesssim-0.3$ at $\delta t\approx2.2$~days), with $\numax\lesssim6$~GHz at $\delta t\approx2.2$~days. Finally, from the NIR F125W observations, we measure a decline rate of $\alpha_{\rm NIR} \lesssim -1.7$ between $\delta t = 3.6$ and $16.4$~days. Next, we use the $\alpha$-$\beta$ closure relations \citep{gs02} to infer the location of the cooling frequency, $\nuc$, relative to the X-ray band. We calculate the value of $p$ from both the spectral and temporal indices of the XRT data for two scenarios: $\numax < \nuX < \nuc$ and $\nuX > \nuc$, requiring the value of $p$ to be in agreement within each scenario. We find consistency between the observed X-ray light curve spectrum and decline rate for $\nuX < \nuc$, with $p=1.90 \pm 0.13$ from $\alpha_{\rm X}$ and $p=1.94 \pm 0.40$ from $\beta_{\rm X}$, with a weighted mean and 1$\sigma$ uncertainty of $\langle p \rangle = 1.90 \pm 0.13$.

\subsection{A Near-Infrared Excess}
\label{sec:optrouble}
We now demonstrate that the NIR observations cannot be reconciled with the X-ray and radio observations in a simple FS model. The shallow radio light curve between $\delta t = 0.23$ and 2.19~days followed by a decline, together with the shallow radio spectral index at $\delta t\approx2.2$~days, suggest that $\numax$ passes through the radio band between the first two radio observations. Taking $\numax\approx6$~GHz at $\delta t \approx1$~day, we require $F_{\nu,\rm max}\approx F_{\nu,{\rm radio}}\approx25~\mu$Jy. At the time of the {\it HST} observations at $\delta t=3.5$~days, we thus expect $\numax\approx0.9$~GHz. For a maximally shallow spectral index of $\beta_{\rm radio-NIR}\approx-0.5$, this gives a predicted NIR flux of $F_{\nu,{\rm F125W}}\approx0.049\,\mu$Jy. Even in this optimistic case, the predicted flux is $\approx10$ fainter than the observed value of $F_{\nu,{\rm F125W}}\approx0.55\,\mu$Jy. 

In fact, the observed spectral index between the predicted radio and observed NIR fluxes at $\delta t = 3.5$~days is extremely shallow, with $\beta_{\rm R-NIR}\approx-0.3$, which cannot be explained in the context of a FS model.
We find that any model which fits the X-ray and radio behavior will under-estimate the observed NIR flux by factors of $\gtrsim 5-10$, and requires a NIR excess. The NIR excess flux, relative to representative afterglow light curve and spectral energy distribution (SED) models are shown in Figure~\ref{fig:aglc}. In this first scenario (Scenario I), we subsequently model the X-ray and radio afterglows with a FS model and address the NIR excess emission separately in Section~\ref{sec:kn}. We present an alternative scenario to explain the entire broad-band data set (Scenario II) in Section~\ref{sec:bb_alternative}.

\subsection{X-ray and Radio Afterglow Modeling}
Setting aside the NIR emission as arising from an additional component, we now outline the available constraints and priors from the radio and X-ray observations, and use Markov Chain Monte Carlo (MCMC) analysis to determine the median values and posteriors in the burst explosion properties. We find that for typical parameters, the self-absorption frequency, $\nua\approx0.8~{\rm GHz}\EKiso^{1/5}\dens^{3/5}<\nu_{\rm R}$. In this regime (the $\nu^{1/3}$ power-law segment), the radio flux density is sensitive to a combination of kinetic energy and circumburst density ($F_{\nu,\rm R}\propto \EKiso^{5/6}\dens^{1/2}$). For the X-ray band, our inference that $\nu_m<\nu_X<\nu_c$ provides an additional constraint on the combination of energy and density ($F_{\nu,X}\propto \EKiso^{(3+p)/4}\dens^{1/2}$). Since the flux density in both observing bands depend on \dens\ in the same way, the density is expected to be very weakly constrained for this burst. 
In this regime, the X-ray and radio observations, together with the constraint that $\nuc>\nuX$, require $\epsb\lesssim6\times10^{-2}$ for $\epse \approx 0.1$ and $p\approx2.05$.

We, therefore, consider two values of $\epsb = 10^{-2}$ and $10^{-3}$, selected to be consistent with the above derived constraint, and also matched to the few values of $\epsilon_B$ that have been derived for short GRBs \citep{fbm+15}, to estimate $\EKiso$ and $\dens$. We follow the methods outlined in \citet{fbm+15}, which uses the afterglow flux densities to map to an allowed parameter space for kinetic energy and density. Using the 6.05~GHz observation at $\delta t = 0.23$~days of $F_{\nu,{\rm R}}=33.4 \pm 8.2\,\mu$Jy, and the first XRT detection at $\delta t = 0.006$~days of $F_{\nu,X}=0.33 \pm 0.08\,\mu$Jy, we determine the respective solutions in the allowed $E_{\rm K,iso}$-$n$ parameter space. Since the radio and X-ray bands are on different spectral segments, they each provide a unique solution. Taking advantage of the fact that $\nu_c > \nu_X$, we also include an upper limit constraint on the location of the cooling frequency assuming a minimum value at the upper edge of the X-ray band, of $\nu_{\rm c, min} = 2.4 \times 10^{18}$~Hz (corresponding to 10~keV). We combine the probability distributions from the two solutions and constraints to obtain a 2D solution, and marginalize over the parameter space to obtain 1D solutions: $\log(\EKiso/{\rm erg})=51.09\pm0.22$ and $\log(\dens/{\rm cm}^{-3})=-1.6\pm0.50$
for $\epsb = 10^{-2}$ and $\log(\EKiso/{\rm erg})=52.06\pm0.24$ and $\log(\dens/{\rm cm}^{-3})=-2.54\pm0.54$ for $\epsilon_B=10^{-3}$.
We use these probability distributions of $\EKiso$ and $\dens$ in our subsequent multi-wavelength modeling as lognormal priors on the corresponding parameters, together with a uniform prior on $p\in[2.001,3.01]$. We fix $\epse=0.1$ \citep{pk02,ss11}, and carry out the modeling using both representative values of $\epsb$. Our priors and assumptions for fixed values are listed in Table~\ref{tab:mcresults}.

\begin{figure}
\centering
\includegraphics[width=\columnwidth]{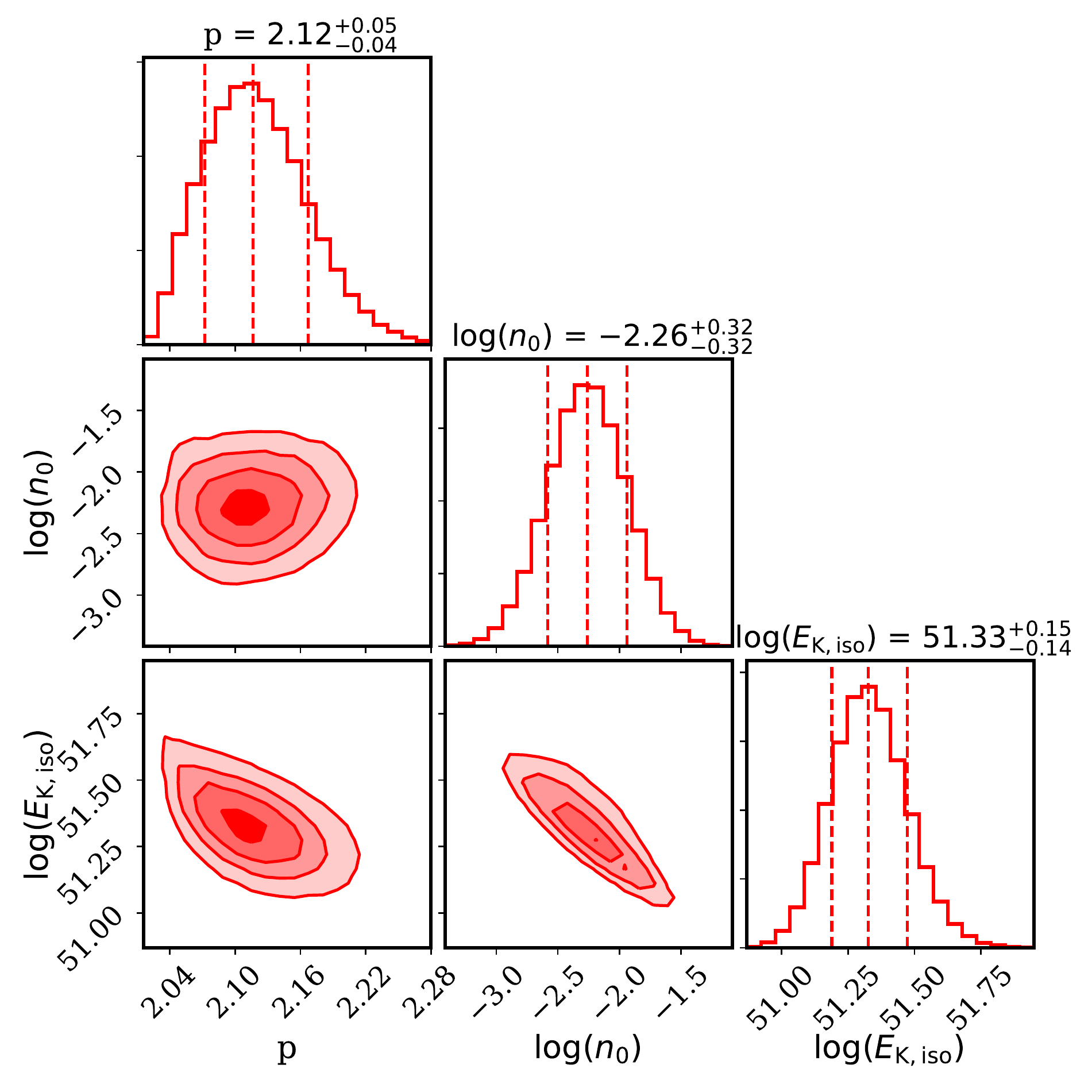}
\vspace{-0.1in}
\caption{Posterior probability density functions and parameter correlations from MCMC fitting for Scenario~I (NIR excess), of all available X-ray and radio afterglow observations of \grb\ for $\epsb=10^{-2}$. 
In each posterior distribution, vertical lines denote the median and 68\% confidence intervals, while contours in the parameter correlation plots correspond to $1-$, $2-$ and $3\sigma$ solutions, respectively. We have fixed $\epse=0.1$ and employed uniform priors on $p\in[2.001,3.01]$. For $\dens$ and $\EKiso$, we used the constraints derived from the radio and X-ray detections as log normal priors (Table~\ref{tab:mcresults}; the derived correlation between these parameters is consistent with the expectation for when $\nuc$ is unconstrained, $\EKiso\propto\dens^{-1/3}$.).
\label{fig:ag_corner}}
\end{figure}

\begin{deluxetable*}{llrrrr}[]
\linespread{1.2}
\tablecaption{GRB\,200522A afterglow parameters \label{tab:mcresults}}
\tablecolumns{7}
\tablewidth{0pt}
\tablehead{
\colhead{Parameter} &
\colhead{Units} &
\multicolumn{2}{c}{Scenario I: FS-only} &
\colhead{} &
\multicolumn{1}{c}{Scenario II: FS+RS}
}
\startdata
$\epse^\dag$ & \nod & 0.1 & 0.1 & & 0.3 \\
$\epsb^\dag$ & \nod & $10^{-2}$ & $10^{-3}$ & & 0.3 \\
$\log(\EKiso)^\ddag$ ({\it prior}) & erg & $51.09 \pm 0.22$ & $52.06\pm0.24$ & & $50.16 \pm 0.22$ \\
$\log(\dens)^\ddag$ ({\it prior}) & cm$^{-3}$ & $-1.6 \pm 0.5$ & $-2.54\pm0.54$ & & $-1.30 \pm 0.21$ \\
$\log(\EKiso)$ (posterior) & erg & $51.33^{+0.15}_{-0.14}$ & $52.17\pm0.17$ & & $50.20^{+0.09}_{-0.07}$ \\
$\log(\dens)$ (posterior) & cm$^{-3}$ & $-2.26\pm0.32$ & $-2.46\pm 0.40$ & & $-1.32\pm 0.18$ \\
$p$         & & $2.12^{+0.05}_{-0.04}$ & $2.05^{+0.03}_{-0.02}$ & &  $2.15^{+0.08}_{-0.05}$ \\
$\tjet$     & days & $>3.5$ & $>3.5$ & & $3.38^{+0.97}_{-0.66}$ \\
$\thetajet$  & deg & $>6.7$ & $>6.3$ & & $14.61^{+1.33}_{-1.13}$\\
$\log(\EK)$  & erg & $48.93-51.33^\star$ & $49.96-52.17^\star$ & & $48.72^{+0.08}_{-0.07}$ \\
\enddata
\tablecomments{Afterglow priors and posteriors for two scenarios: (I) a spherical, forward shock model to fit the radio and X-ray bands, leaving a NIR excess and (II) a joint forward and reverse shock model with a jet break to explain the broad-band data set. \\
${}^\dag$ Fixed parameters. \\
${}^\ddag$ Derived from preliminary considerations, and used as priors for the MCMC. \\
$^\star$ Lower limit is set by the constraint on the jet opening angle, while the upper limit is set by the isotropic-equivalent value.}
\end{deluxetable*}

\subsection{Markov Chain Monte Carlo}
\label{sec:mcmc_nirexcess}
We now explore the parameter space of $\dens$, $\EKiso$, and $p$ in this scenario, using the modeling framework described in \cite{lbt+14}.
We incorporate upper limits into the log-likelihood assuming a Gaussian error function. We run 10000 MCMC iterations, discarding the first few steps as burn-in, after which the log-likelihood and parameter distributions appear stationary. 
We thin the output samples by a factor of 10, and plot correlation contours and histograms of the results in Figure~\ref{fig:ag_corner}. 
We list the median parameters derived from the MCMC fit for both values of $\epsb$ in Table~\ref{tab:mcresults}. As expected, the energy and density are poorly constrained, and the output posterior is very similar to the input priors. We do, however, probe the joint density between the two parameters, and find that the major axis of the correlation is aligned along the direction given by $\EKiso\propto\dens^{-1/3}$. This relation is consistent with the expected degeneracy when \nuc\ is unknown \citep{lbt+14}, indicating that $\nuc>\nuX$ provides the dominant source of the correlation between these parameters.

We plot synchrotron light curves for a representative model in Figure~\ref{fig:aglc}. 
For the median parameters, we calculate $Y_{\rm max}\approx0.3$ and $Y_{\rm max}\approx0.7$ (at $\delta t\approx0.1$~days) for $\epsb=10^{-2}$ and $\epsb=10^{-3}$, respectively, confirming that IC cooling occurs deep in the KN regime and does not modify the synchrotron cooling frequency significantly, thus validating our previous assumption regarding IC cooling.

We find that the median parameters of $\EKiso\approx (2-20)\times 10^{51}$~erg and $n=(3.4-5.5) \times 10^{-3}$~cm$^{-3}$ are close to the median values of cosmological short GRBs for the same values of $\epsilon_B$ \citep{fbm+15}. Furthermore, the X-ray and radio data constrain the time of any potential jet break due to collimation effects to $\delta t \gtrsim 3.5$~days, translating to $\theta_{\rm jet} \gtrsim 6.5^{\circ}$ for the median values of the \EKiso\ and \dens\ \citep{spn98}, comparable to some limits measured for short GRBs. Finally, using the derived range of $\EKiso$, and the value of $E_{\gamma,{\rm iso}}$15-150~keV)$\approx 8.4\times 10^{49}$~erg derived in Section~\ref{sec:disc}, we calculate a gamma-ray efficiency of $\eta \approx 0.04$.

\begin{figure*}
\centering
 \begin{minipage}{\textwidth}
   \centering
   $\vcenter{\hbox{\includegraphics[width=0.49\textwidth]{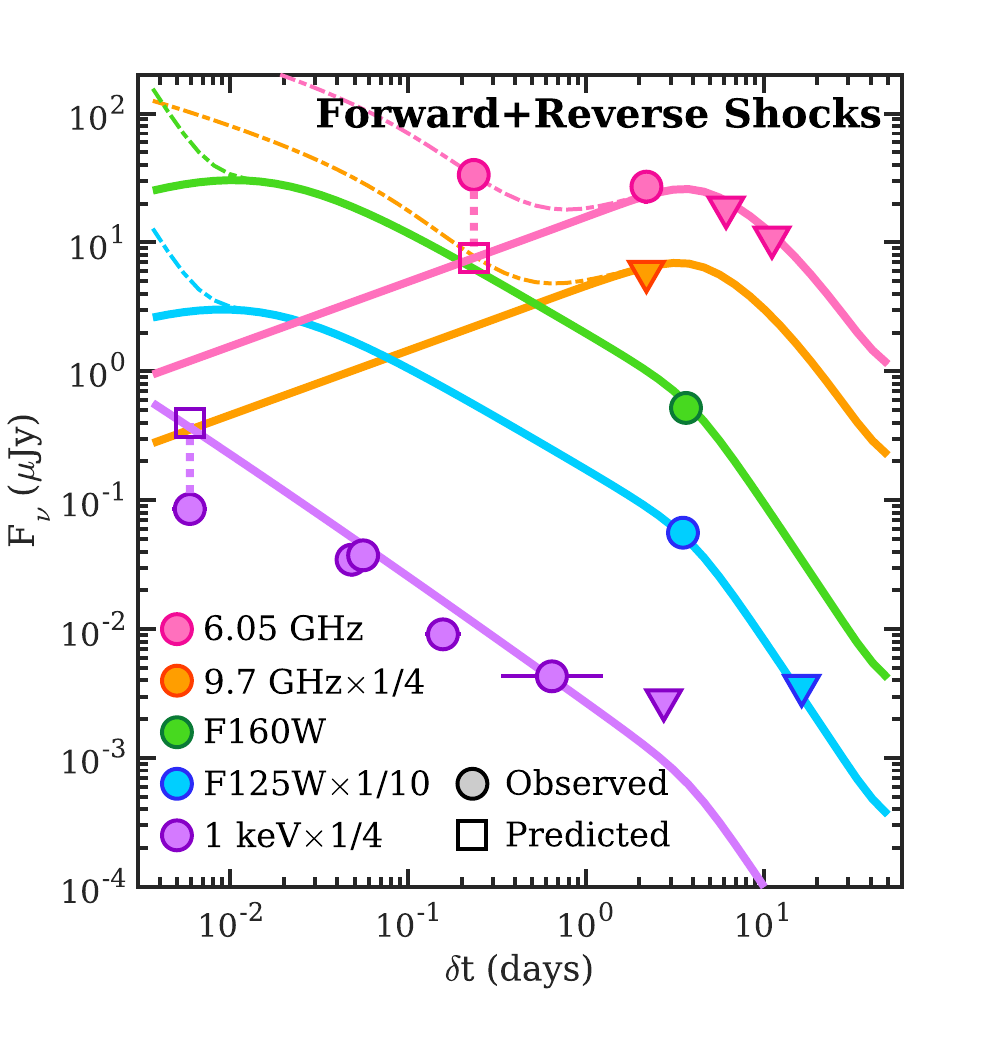}}}$
   $\vcenter{\hbox{\includegraphics[width=0.49\textwidth]{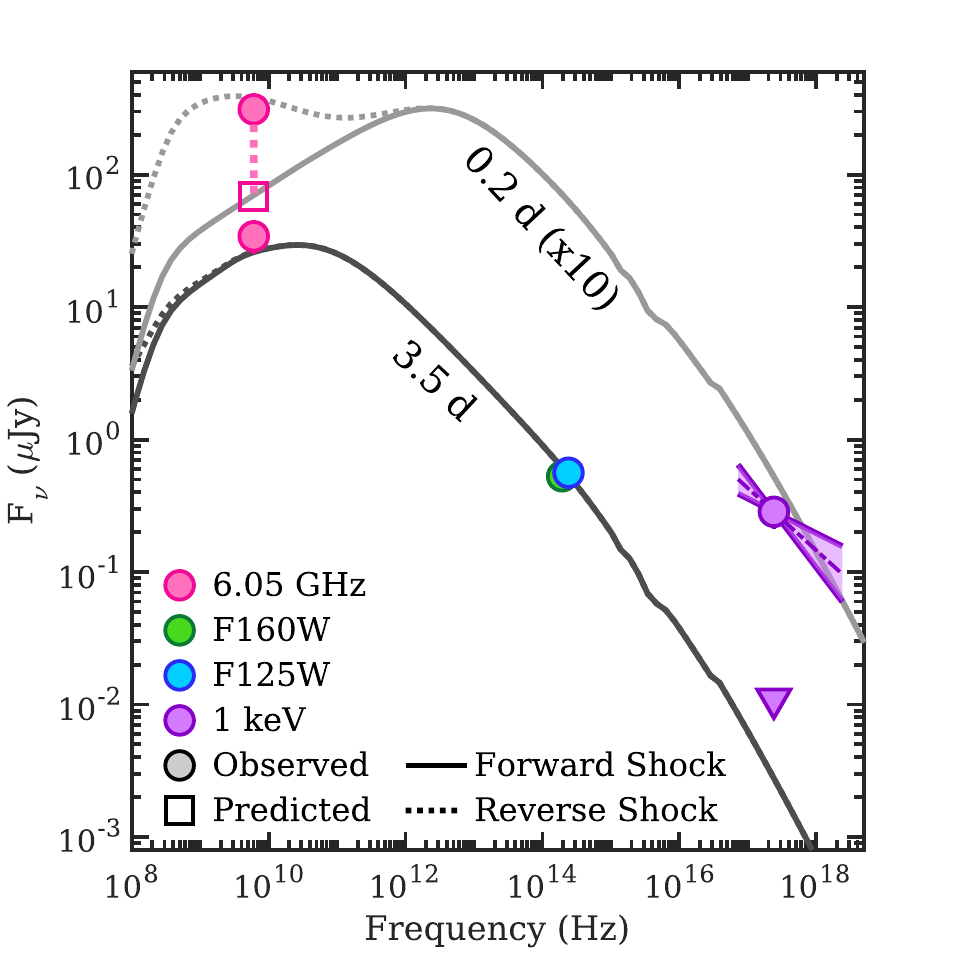}}}$
 \end{minipage}
 \vspace{-0.2in}
\caption{The radio, NIR and X-ray observations of the counterpart of \grb\ (circular points) and models in Scenario II. {\it Left:} Representative afterglow model light curves representing a forward shock with an achromatic jet break at $t_{\rm jet}=4.0$~days (solid lines). The radio data point at $\delta t \approx 0.23$~days is in excess of the model, and can be explained by the addition of a reverse shock (dot-dashed lines). {\it Right:} The corresponding afterglow model spectral energy distributions at $\delta t = 0.2$~days and $3.5$~days, including forward shock only (solid lines) and forward and reverse shocks (dot-dashed lines). In this scenario, the NIR-band temporal evolution is consistent with the forward shock model with a jet break, but is steeper than the observed X-rays, and under-predicts the early radio emission. In addition, the measured X-ray spectral slope (purple regions, representing $1\sigma$ confidence region) is shallower than the predicted slope of $\beta_X=-1$. In both panels, models and data points are scaled as denoted for clarity. Error bars correspond to $1\sigma$ and are generally smaller than the size of the symbols, and triangles correspond to $3\sigma$ upper limits.
\label{fig:aglc_rs}}
\end{figure*}

\section{Broad-band Modeling II: A Collimated Outflow with a Reverse Shock}
\label{sec:bb_alternative}

We can alternatively ameliorate the inconsistency between the radio, NIR, and X-ray observations outlined in Section~\ref{sec:optrouble} by not requiring the FS to explain the first radio detection at $\approx0.2$~days. If we extend the $\nu^{1/3}$ segment to $\gtrsim6$~GHz by increasing $\numax$, the resultant spectrum above $\numax$ can be made to pass through the NIR detection. 
Now, since $\alpha_{\rm R}=0.5$ for $\nua<\nu_{\rm R}<\numax$, we would expect $F_{\rm radio,FS}(0.2~\rm days)\approx9~\mu$Jy, which is a factor of $\approx3$ fainter than the observations. Therefore, we must explain the first radio detection by another component in this model. Early excess flux at radio bands has sometimes been attributed to reverse shock (RS) emission both in long and short GRBs \citep{kfs+99,sbk+06,lbz+13,ltl+19,tcb+19}.
Owing to the limited information available, a variety of RS models are possible. We label this set of models Scenario II.

\subsection{Preliminary considerations}
To derive constraints on the physical parameters in this scenario, we first compare the observed X-ray and NIR behavior to expectations in a standard FS model, as any RS is not expected to contribute significantly in these bands at the times of our observations. The X-ray flux density, extrapolated as a single power law to the time of the first {\it HST} observations at $\delta t\approx3.5$~days, is $F_{\rm \nu,X}\approx0.0057\,\mu$Jy. Relative to the observed value of $F_{\nu,{\rm F125W}}\approx0.55\,\mu$Jy, this yields a NIR-to-X-ray spectral index of $\beta_{\rm NIR-X}=-0.66\pm0.06$, significantly steeper than the measured $\beta_{\rm X}\approx-0.47$.
Therefore, simply extending the FS emission as a single $\beta\approx-0.5$ power law past the NIR would over-predict the X-ray flux by a factor of $\approx5$ unless an additional spectral break were to be present between the NIR and X-ray bands.
If we identify this break as $\nuc$, we expect an X-ray spectral index of $\beta_{\rm X}\approx-1$ and a light curve decline rate of $\alpha_{\rm X}\approx-1$. The former is steeper than the observed value of $\beta_{\rm X}=-0.47^{+0.24}_{-0.19}$, and the latter is steeper than the observed value of $\alpha_{\rm X}\approx-0.67$. The shallow X-ray spectrum cannot be easily reconciled, and remains a concern for any model attempting to explain the X-ray and NIR observations as arising from a synchrotron FS emission.

On the other hand, we note that fitting the X-ray light curve at $\delta t\gtrsim4\times10^{-2}$~days yields a steeper power law, $\alpha_{\rm X}=-0.85\pm0.15$, than that obtained from fitting the entire X-ray light curve, and that this latter value is consistent with the expected decline of $\alpha\approx-1$ for the regime $\numax<\nu_{\rm NIR}<\nuc<\nuX$. Naturally, extrapolating this slope back in time over-predicts the first X-ray detection at $\delta t\approx6\times10^{-3}$~days, which is one of the shortcomings of this model. One possible solution to this is a continuous injection of energy into the FS at $6\times10^{-3}$--$5\times10^{-2}$~days, such that the FS energy increases by a factor of $\approx4$ during this period. Similar injection episodes have been inferred for long-duration GRBs in the past \citep{rm98,bgj04,lbm+15}. A similar effective energy injection could also be attributable to a slightly off-axis viewing geometry of the jet core at $\lesssim4\times10^{-2}$~days. However, given the paucity of data, it is not possible to obtain meaningful constraints on the either effect and we, therefore, do not attempt it here. We ignore the first X-ray data point at $6\times10^{-3}$~days in our subsequent analysis under Scenario II. 

\subsection{Jet break}
In this scenario, the NIR detection at $\delta t\approx3.5$~days arises from FS synchrotron emission in the regime $\numax<\nu_{\rm NIR}<\nuc$. From the X-ray light curve, we have inferred that $p\approx2$. This implies an NIR decay rate of $\alpha_{\rm NIR}\approx-0.75$. However, the F125W upper limit at $\delta t\approx16.4$~days implies a much steeper decline of $\alpha_{\rm NIR}<-1.8$ at $\delta t\gtrsim3.5$~days. 

GRB jets are expected to be collimated outflows, and the signature of ejecta collimation has previously observed in short GRB light curves \citep{nfp09,fbm+15}. One possibility that could explain the steep NIR light curve is that a jet break occurs at $3.5\lesssim t_{\rm jet}\lesssim16.4$~days, and we include the possibility of a jet break in our MCMC modeling within Scenario II in the next section.

\begin{figure*}
\centering
\includegraphics[width=0.8\textwidth]{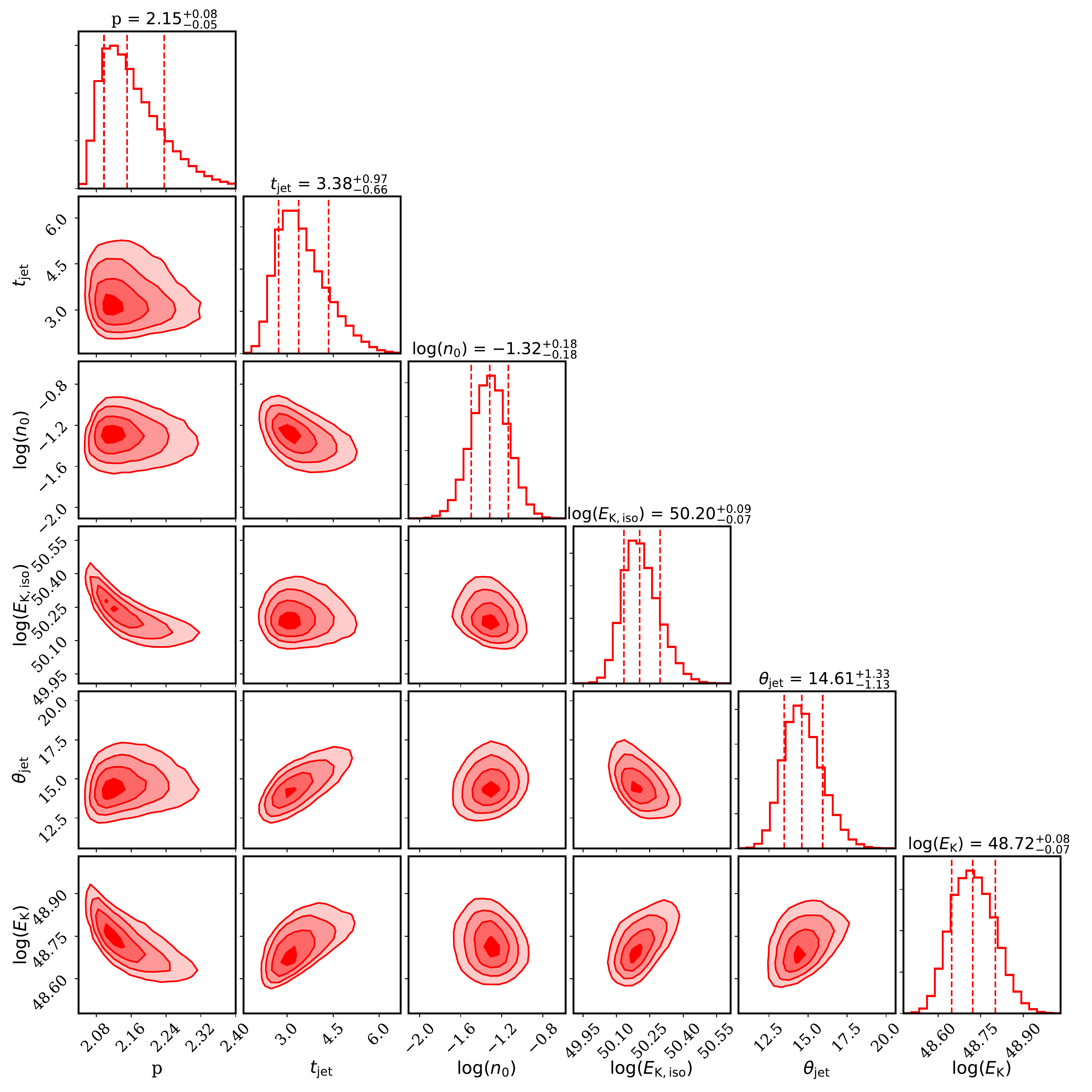}
\vspace{-0.1in}
\caption{{\it} Posterior probability density functions and parameter correlations from Markov Chain Monte Carlo fitting for Scenario~II (RS + jet break) of all available afterglow observations of \grb, ignoring the first radio and X-ray detections (i.e., in the RS model).
In each posterior distribution, vertical lines denote the median and 68\% confidence intervals, while contours in the parameter correlation plots correspond to $1-$, $2-$ and $3\sigma$ solutions, respectively. We have fixed $\epse=\epsb=0.3$ and employed uniform priors on $p\in[2.001,3.01]$ and and $(t_{\rm jet}/{\rm days})\in[10^{-5},10^5]$. 
For $\dens$ and $\EKiso$, we used the constraints derived from the radio, NIR, and X-ray observations as lognormal priors (Table~\ref{tab:mcresults}). The opening angle ($\theta_{\rm jet}$ in degrees) and the beaming-corrected kinetic energy ($E_{\rm K}$ in erg) are derived from the individual Monte Carlo samples. 
\label{fig:ag_corner_rs}}
\end{figure*}

\subsection{Markov Chain Monte Carlo}
\label{sec:mcmc_rs}
We now consider constraints imposed upon the physical parameters by this RS+FS model. Requiring $\nu_{\rm opt}<\nuc<\nuX$, taking $p\approx2.05$ and matching the observed radio flux density at $\approx2.2$~days and the X-ray flux density at $\approx0.05$~days, we find that no solutions are possible for $\epsb<1$, unless $\epse\gtrsim0.3$. Taking $\epse\approx0.3$, we find $\epsb\gtrsim0.3$, $\dens\gtrsim2\times10^{-2}$, and $\EKiso\lesssim2\times10^{50}$~erg. 
Once again following the methods of \citet{fbm+15} in the regime $\nua<\nu_{\rm R}<\numax<\nu_{\rm opt}<\nuc<\nuX$ and including the constraint $\nu_c < \nu_X$, we obtain $\log(\EKiso/{\rm erg})=50.16\pm0.22$ and $\log(\dens/{\rm cm}^{-3})=-1.30\pm0.21$. We use these probability distributions of $\EKiso$ and $\dens$ in multi-wavelength modeling as lognormal priors on the corresponding parameters. We fix $\epse=\epsb=0.3$, and leave $p$ and $\tjet$ as additional free parameters. 

We do not include the radio point at $\approx0.2$~days (dominated by the RS in this scenario) and the first X-ray point at $\approx6\times10^{-3}$~days (as this cannot be explained in this model). We run and process MCMC iterations in a similar fashion as for Scenario~I. We plot a representative model from our fits in Figure~\ref{fig:aglc_rs}. 
Since $\nuX<\nuc$ in this scenario, the X-ray band is sensitive to $\EKiso$, and so this parameter (and, consequently, also $\dens$) is slightly better constrained than in Scenario I. Interpreting the NIR steepening as a jet break allows us to constrain $\tjet\approx3.4$~days, around the time of the NIR detection, which yields a fairly wide opening angle of $\approx14^{\circ}$. We follow \cite{sph99} to calculate \thetajet\ from \tjet, \EKiso, and \dens, and calculate the beaming-corrected kinetic energy (\EK) for each sample. We plot correlation contours between the parameters from the fit in Figure~\ref{fig:ag_corner_rs} and list summary statistics from the marginalized posterior density functions in Table~\ref{tab:mcresults}. 

In this interpretation, there is only one detection of the putative RS, and thus it is impossible to constrain its properties fully. Under the assumption that $\nu_{\rm a,RS}<\nu_{\rm m,RS}\lesssim\nu_{\rm R}$ at $\approx0.2$~days, we require 
$F_{\nu,\rm m,RS}\approx80~\mu{\rm Jy}(\nu_{\rm m,RS}/{\rm GHz})^{-0.5}(t/0.2~{\rm days})^{-1.5}$, where we have assumed a spectral index of $\approx(1-p)/2\approx-0.5$ above the RS peak and the time evolution of $\nu_{\rm m,RS}$ is appropriate for either a relativistic RS (where it is expected to evolve as $t^{-73/48}$;  \citealt{kob00}) and for a non-relativistic RS for the $g$-parameter, $g\approx2.2$ (where it is expected to evolve as $t^{-(15g+24)/(14g+7)}$\citealt{ks00}). For our representative FS model in Scenario II, we have $F_{\nu, \rm m,FS}\approx80~\mu$Jy. Thus, the initial Lorentz factor (assuming equal magnetization of the FS and RS), $\Gamma_0\approx F_{\nu,\rm m,RS}(t_{\rm dec})/F_{\nu,\rm m, FS}$, where $t_{\rm dec}$ is the deceleration time \citep{ks00}. This yields,
\begin{equation}
    \Gamma_0 \approx \left[\frac{\nu_{\rm m,RS}(0.2~\rm days)}{\rm GHz}\right]^{-0.5}
    \left[\frac{t_{\rm dec}}{0.2~\rm day}\right]^{-1.5}.
\end{equation}
Taking $t_{\rm dec}\lesssim6\times10^{-3}$~days, the time of the first X-ray detection, and $\nu_{\rm m,RS}\lesssim6$~GHz at 0.2~days, we find a reasonable value for the initial ejecta Lorentz factor, $\Gamma_0\gtrsim80$. We include one such RS model in Figure~\ref{fig:aglc_rs}.

\section{Host Galaxy and Environmental Properties}
\label{sec:host}

\subsection{Stellar population modeling}
\label{sec:hostsps}

Using the Pan-STARRS1 Source Types and Redshifts with Machine learning (PS1-STRM) catalog \citep{bsf+19}, the next two closest catalogued galaxies besides the host of \grb\ have redshifts of $z_{\rm phot}=0.89$ and $z_{\rm phot}=0.55$ at $\delta R=10.3''$ and $11.9''$, respectively. While the nearby galaxy at a similar redshift of $z_{\rm phot} \approx 0.55$ could point to an origin in a group, given the star-forming nature of the host coupled with the fairly even photometric redshift distribution of surrounding galaxies, it is unlikely that this burst is part of a low-redshift galaxy cluster.

We model the stellar population properties of the host galaxy of \grb\ with \texttt{Prospector}, a Python-based stellar population inference code \citep{Leja_2017}. We use \texttt{Prospector} to determine the following stellar population properties and characteristics: stellar mass ($M_*$), mass-weighted stellar population age ($t_m$), dust attenuation ($A_V$), stellar metallicity ($Z_*$), and star formation history (SFH) characterized by an $e$-folding factor $\tau$. We apply a nested sampling routine with \texttt{dynesty} \citep{Dynesty} to the observed photometry and spectroscopy and produce model SEDs with \texttt{Python-fsps} (Flexible Stellar Population Synthesis; \citealt{FSPS_2009, FSPS_2010}). For our fits, we fix redshift to the value of the spectroscopically-determined redshift, $z=0.5536$ (see Section \ref{sec:hostobs}) and leave all other parameters free. We jointly fit the observed photometry and spectrum of the host of \grb\ weighted by the $1\sigma$ photometric uncertainties and error spectrum.

We initialize our stellar population models with a Chabrier initial mass function (IMF; \citealt{2003PASP..115..763C}) and Milky Way Dust Extinction Law \citep{MilkyWay}. We use a parametric, delayed-$\tau$ star formation history (SFH), given by: 
\begin{equation}
    \text{SFR}(t) = M_F\times \left[\int_0^t{te^{-t/\tau} dt}\right]^{-1} \times t e^{-t/\tau},
\end{equation}
where SFR is star formation rate, $M_F$ is the total mass formed from dust to stars over the lifetime of the galaxy, and $t$ represents the age of the galaxy at which star formation commences. {\tt Prospector} provides posteriors on $M_F$, $t$, and $\tau$ from which we determine the posteriors in $M_*$ and mass-weighted age, $t_m$, using the SFH and analytic conversions from total mass to stellar mass (Equation~2 in \citealt{Leja2013}, and detailed on \citealt{nfd+20}). We choose $t_m$ as the stellar population age metric, to avoid disproportionately weighting contributions from younger, brighter stars (as is the case for simple stellar population ages; \citealt{Conroy2013SED}) and to provide a more robust estimate of when the short GRB progenitor could have formed.

We also employ a $10^{\text{th}}$-order Chebyshev polynomial to fit the spectral continuum. We include a model for nebular emission, characterized by two additional free parameters: $\log(Z_\text{gas}/Z_\odot)$ which measure gas metallicity and a parameter for gas-ionization. Finally, we impose a 2:1 ratio on the amount of dust attenuation between the younger and older stellar populations, respectively, as young stars in SF regions typically experience twice the amount of dust attenuation as older stars \citep{cab+20, pkb+14}.

\begin{figure*}
\centering
\includegraphics[width=0.9\textwidth]{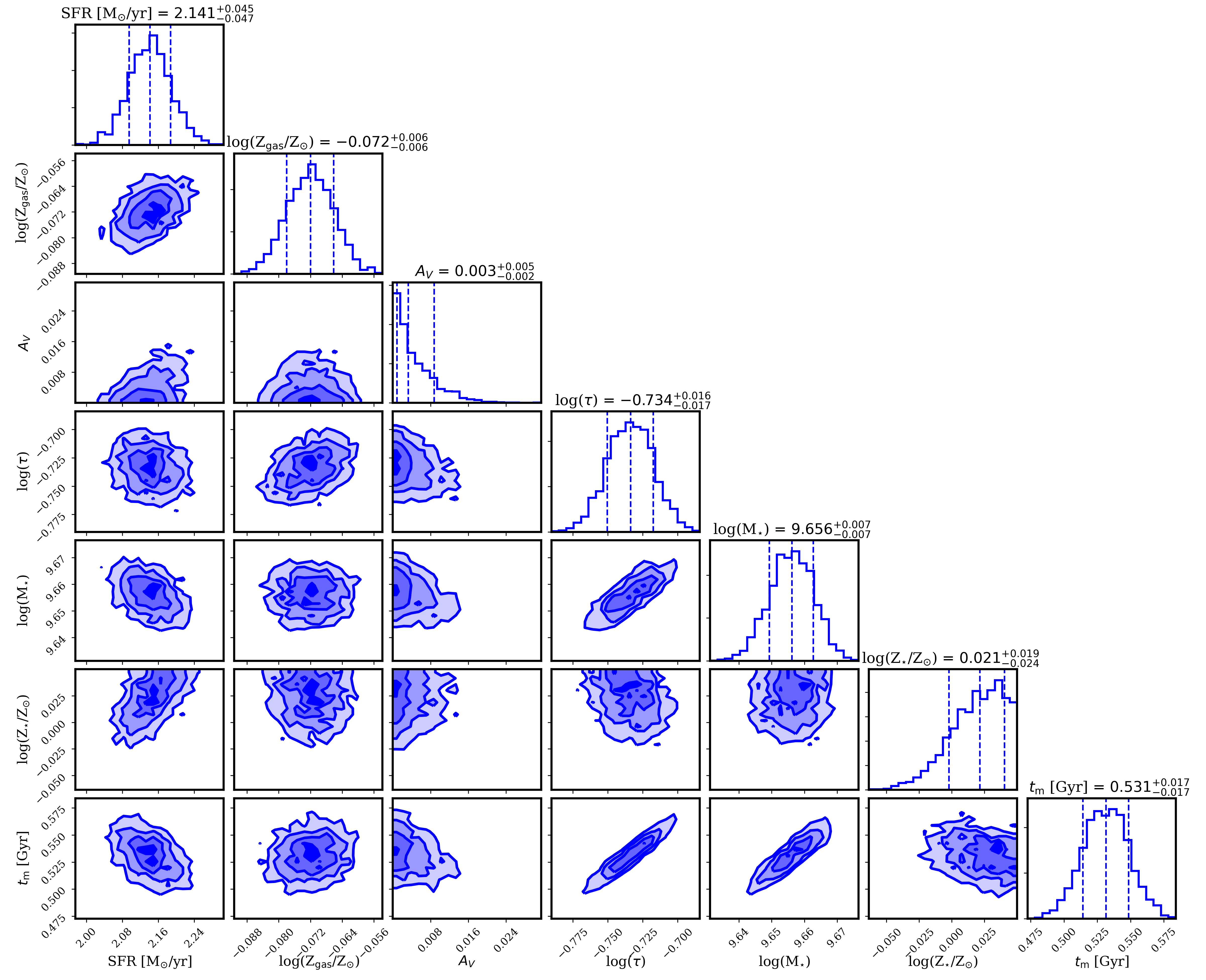}
\includegraphics[width=0.6\textwidth]{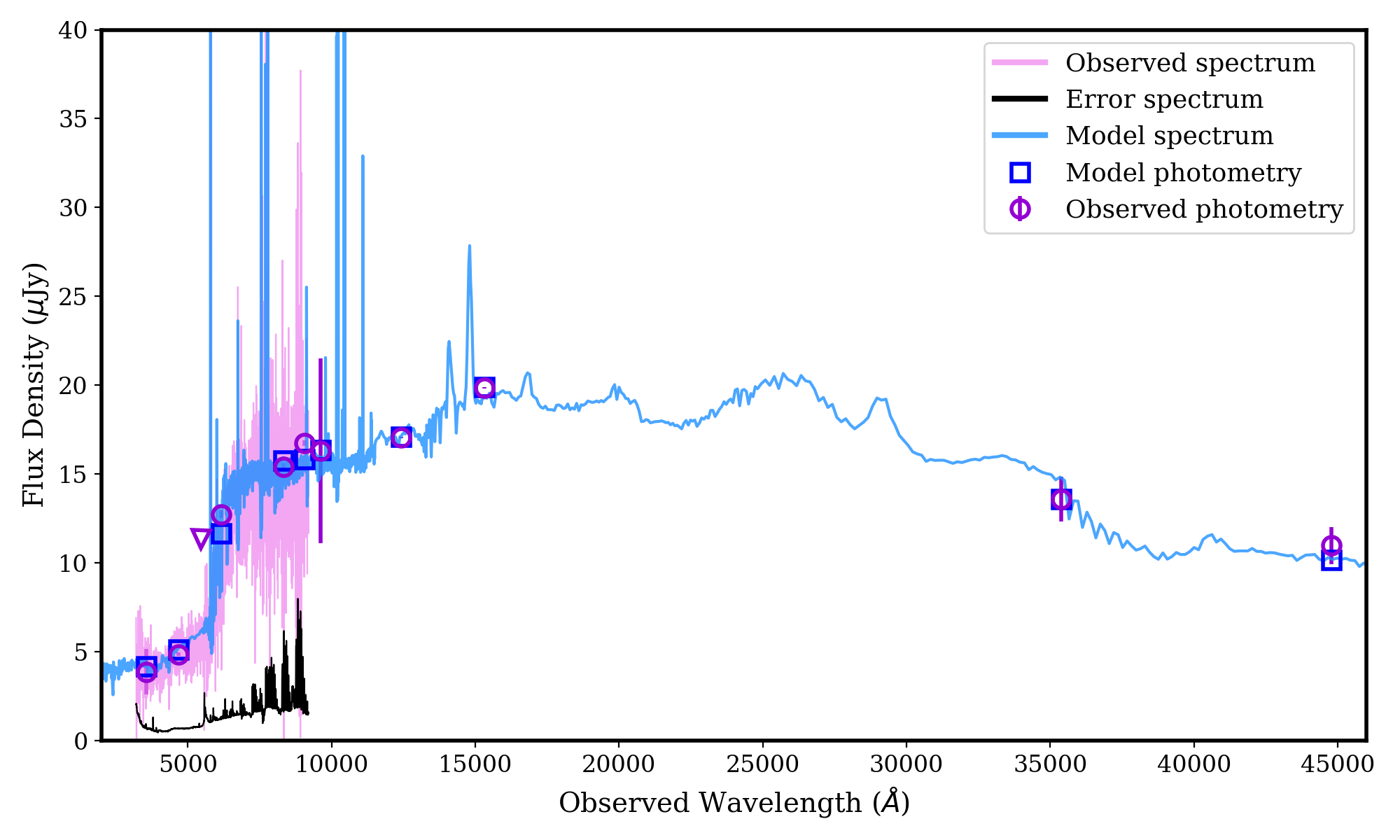}
\vspace{-0.1in}
\caption{{\it Top:} Posterior distributions and parameter correlations from joint fitting of the spectrum and multi-band photometry of \grb\ with {\tt Prospector}. In each posterior distribution, vertical lines denote the median and 68\% confidence intervals, while contours in the parameter correlation plots correspond to $1-$, $2-$ and $3\sigma$ solutions, respectively. {\it Bottom:} Keck/LRIS spectrum of the host galaxy of \grb\ (light pink) along with the $uGVRIZy$-band, F125W, F160W, and 3.6 and $4.5\,\mu$m photometry from SDSS DR12 \citep{aaa+15}, Pan-STARRS1 \citep{cmm+16}, Keck/LRIS, Keck/DEIMOS, HST/WFC3, and Spitzer \citep{Timlin2016,Papovich2016} (pink circles/triangle). The model spectrum and photometry characterized by the median values for the stellar population properties are also shown (blue line and squares, respectively). Overall, the model matches the continuum of the observed spectrum, the strength of the $4000$\AA\ break, the photometric colors, and the locations of the nebular emission lines. 
\label{fig:corner}}
\end{figure*}

We present the resulting posterior distributions of the free parameters in Figure \ref{fig:corner} and report the median values and bounds corresponding to 68\% credible intervals in Table \ref{tab:prop}. The observed host galaxy photometry and spectrum, along with the model spectrum and photometry characterized by the {\tt Prospector} median parameters, is shown in Figure~\ref{fig:corner}. The shape of the spectrum as well as the locations of the emission lines are well fit by the model. We find that the host is characterized by a young stellar population with $t_m \approx 0.53$~Gyr, $M_* \approx 4.5 \times 10^9 M_\odot$, $A_V \approx 0$, and near-solar stellar metallicity of $\log(Z_*/Z_\odot) \approx 0.02$. The determined $\log(Z_\text{gas}/Z_\odot)$ is $\approx -0.07$, approximately the expected value from the $M-Z$ relation at redshifts of $0.07 < z < 0.7$ \citep{Savaglio2005, Kewley2008}. Based on these parameters, we calculate a SFR of $\approx 2.1 M_\odot$~yr$^{-1}$ and a specific SFR per unit mass (sSFR) of $4.7 \times 10^{-10}$ yr$^{-1}$.

\begin{deluxetable}{ccc}[t!]
\linespread{1.2}
\tablecaption{GRB\,200522A Derived Host Galaxy Properties \label{tab:prop}}
\tablecolumns{3}
\tablewidth{0pt}
\tablehead{
\colhead{Property} &
\colhead{Value} & 
\colhead{Units}
}
\startdata
$z$ & $0.5536 \pm 0.0003$ & \nod \\
t$_m$ & $0.531 \pm 0.017$ & Gyr \\
A$_V$ & $0.003 ^{+0.005}_{-0.002}$ & AB mag\\
log(${\tau}$) & $-0.734 ^{+0.016}_{-0.017}$ \\
log(Z$_{\rm{gas}}$/Z$_{\odot}$) & $-0.072 \pm 0.006$ \\
log(Z$_*$/Z$_{\odot}$) & $0.021 ^{+0.019}_{-0.024}$ \\
log(M$_*$/M$_{\odot}$) & $9.656 \pm 0.007$ \\
SFR (SED) & $2.141^{+0.045}_{-0.047}$ & $M_{\odot}$~yr$^{-1}$ \\
SFR (H$\alpha$) & $4.90 \pm 0.47$ & $M_{\odot}$~yr$^{-1}$ \\
sSFR$^\dagger$ & $4.7-10.5$ & $10^{-10}$~yr$^{-1}$ \\
$r_e$ & 3.9 & kpc \\
$\delta R$ (F125W) & $1.01 \pm 0.35$ & kpc \\
$\delta R$ (F160W) & $0.93 \pm 0.19$  & kpc \\
$\delta R$ (VLA) & $3.44 \pm 2.34$  & kpc \\
$\delta R$ & $0.24 \pm 0.04$  & $r_e$ \\
Frac. Flux (F125W) & 0.95 \\
Frac. Flux (F160W) & 0.96
\enddata
\tablecomments{Properties of \grb\ and its host galaxy determined in this work.\\
$^\dagger$ The range is set by the $H\alpha$ and SED-derived SFRs.}
\end{deluxetable}

\subsection{Nebular Emission Lines}

\begin{deluxetable}{ccc}[t!]
\linespread{1.2}
\tablecaption{GRB\,200522A Emission Line Fluxes \label{tab:emline}}
\tablecolumns{3}
\tablewidth{0pt}
\tablehead{
\colhead{Line} &
\colhead{$\lambda_{\rm obs}$} & 
\colhead{$f$} \\
\colhead{} &
\colhead{\AA} & 
\colhead{($10^{-16}$~erg~s$^{-1}$~cm$^{-2}$)}
}
\startdata
[OII]$\lambda 3727$ & 5791.88 & $5.46 \pm 0.57$\\
H$\gamma$ & 6742.6 & $0.53 \pm 0.43$\\
H$\beta$ & 7552.48 & $1.67 \pm 0.51$\\
${\rm[OIII]}\lambda 4959$ & 7703.71 & $1.07 \pm 0.45$\\
${\rm [OIII]}\lambda 5007$ & 7778.6 & $2.80 \pm 0.49$\\
H$\alpha$ & 10195.88 & $4.81 \pm 0.46$\\
\enddata
\tablecomments{Emission line centroids and integrated line fluxes. Measurements are corrected for Galactic extinction in the direction of the burst.}
\end{deluxetable}

\label{sec:hostemlines}
We measure the flux-weighted centroids and integrated fluxes of the nebular emission lines using a custom Python routine\footnote{\url{https://github.com/CIERA-Transients/MODS_spectroscopy/blob/master/spec\_SFR_metallicity.ipynb}}. The derived line centroids, and emission line fluxes and uncertainties are shown in Table~\ref{tab:emline}. The observed $H\alpha$ to $H\beta$ line ratio of $\approx2.88$ is consistent with the expectation for ionization equilibrium under Case B recombination at a typical nebular temperature of $10^4$~K and electron density of $10^{2}~{\rm cm}^{-3}$ \citep{ost89}. This indicates no additional extinction ($A_{\rm V}\lesssim0.1$~mag) along the line of sight to star-forming regions within the host, consistent with the results from SED fitting. For the observed H$\alpha$ line flux (Table~\ref{tab:emline}), we obtain an H$\alpha$ line luminosity of L(H$\alpha$) = $(6.21 \pm 0.59) \times 10^{41}$ erg s$^{-1}$. Using standard conversions \citep{ken98,mkt+06}, we determine SFR (H$\alpha$) = $4.90 \pm 0.47$ ~$M_{\odot}$~yr$^{-1}$. This is a factor of $\approx 2$ larger than the SED-derived SFR, although we note that both diagnostics can have systematic uncertainties by factors of $\approx 2$ or more \citep{mkt+06,Theios2019}, and we report both values for completeness. The H$\alpha$-derived value gives sSFR$\approx 10.5 \times 10^{-10}$ yr$^{-1}$

Using the calibration of \cite{ccm+17}, searching over a grid of the metallicities derived from the $R_2$, $R_3$, $R_{23}$, and $O_{32}$ metallicity diagnostics (equally weighted), and using the solar photospheric oxygen abundance from \cite{ags+09}, we find a gas-phase metallicity of $12+\log({\rm O/H})=8.54\pm0.03$, or $\log(Z_\text{gas}/Z_\odot)=-0.16\pm0.03$, similar to the value of  $\log(Z_\text{gas}/Z_\odot) \approx-0.1$ from SED modeling.

\subsection{Host Morphology and Fractional Flux}

We use the {\tt GALFIT} software \citep{phi+07} to fit the 2D surface brightness profile of the host galaxy of \grb\ in each of the F125W and F160W images. For each image, we perform a three-component fit representing the galaxy, the neighboring galaxy to the southeast, and the sky background. We use S\'{e}rsic surface brightness profile models for the two galaxies, allowing the centroid, central surface brightness, effective radius ($r_e$) and S\'{e}rsic index $n$ to vary. The resulting best-fit F160W solution is characterized by $n=2.3$ and $r_e=0.60''$ for the host, with $\chi^2_{\nu}=2.2$. For F125W, the best-fit solution is $n=2.1$ and $r_e=0.60''$. At the redshift of \grb, the host effective radius becomes $r_e=3.90$~kpc. Taking into account the size of the host galaxy, we also calculate a host-normalized offset of $\delta R = 0.24 \pm 0.04~r_e$ (Table~\ref{tab:prop}). 

The residual images exhibit a clean subtraction of the neighboring galaxy, an indication that it is well-modeled by {\tt GALFIT}. On the other hand, the residuals for the host galaxy exhibit clear structure in both filters, extending from NW to SE. The galaxy appears to be bulge-dominated with a disturbed outer stellar halo, potentially indicative of a fairly recent galaxy merger or interaction with a neighboring galaxy.

We also determine the location of \grb\ with respect to its host light distribution, using the ``fractional flux'' diagnostic (FF; \citealt{fls+06}). The FF is defined as the fraction of cumulative host light in pixels fainter than brightness level at the counterpart position. It is a complementary diagnostic to probe the burst's location relative to its host galaxy that is independent of host morphology. Using the position of the NIR counterpart, and employing a $1\sigma$ cut-off to determine the bounds of the host galaxy, we calculate fractional flux values of $0.95-0.96$ for the two filters, indicative of a strong correlation with its host stellar mass distribution. The derived morphological properties, offset, and FF values are listed in Table~\ref{tab:prop}.

\section{The Near-Infrared Counterpart of \grb}
\label{sec:kn}

\begin{figure*}
\centering
\includegraphics[width=0.495\textwidth,trim={0cm 0cm 0.5cm 0cm}]{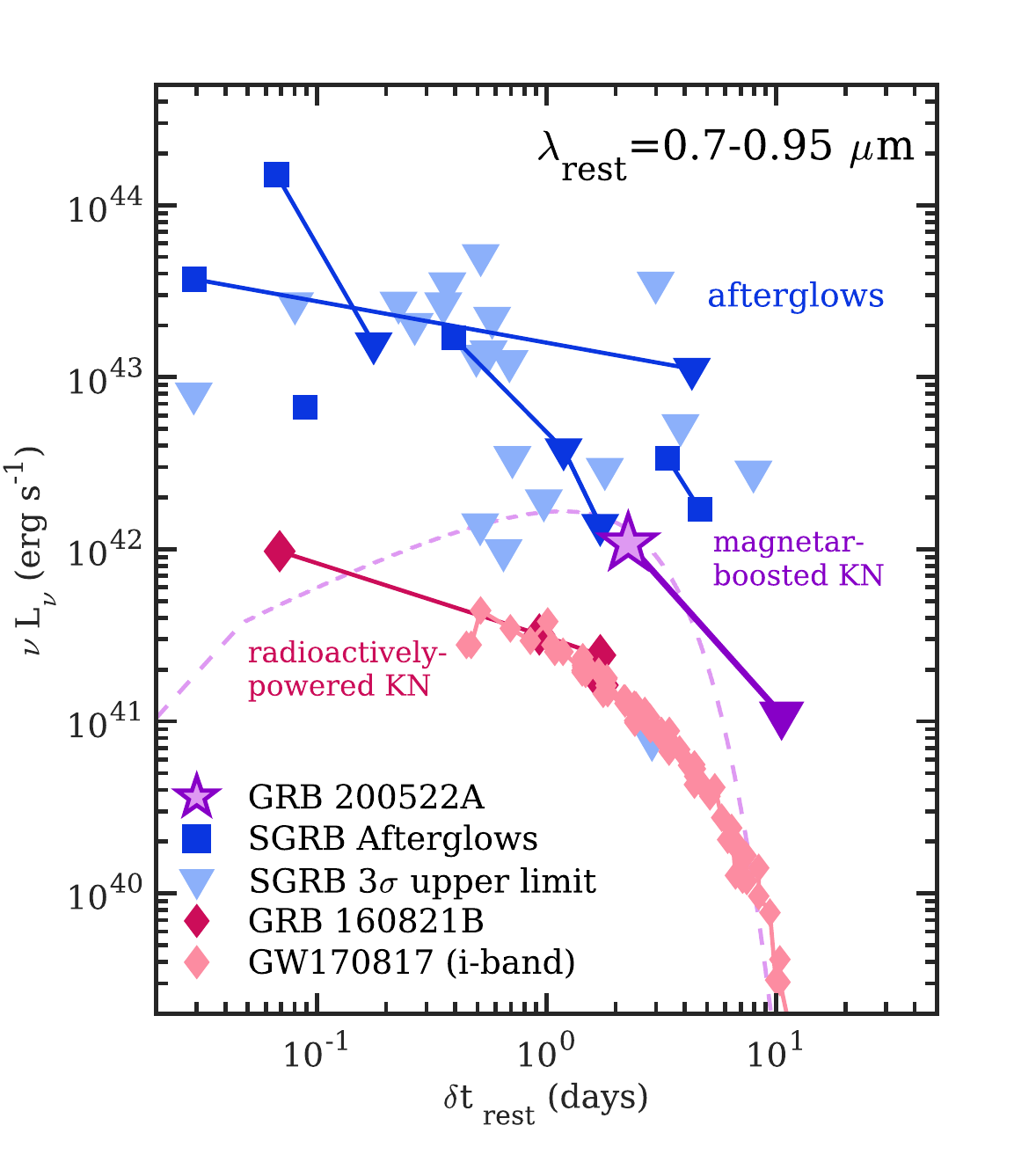}
\includegraphics[width=0.495\textwidth,trim={0.5cm 0cm 0cm 0cm}]{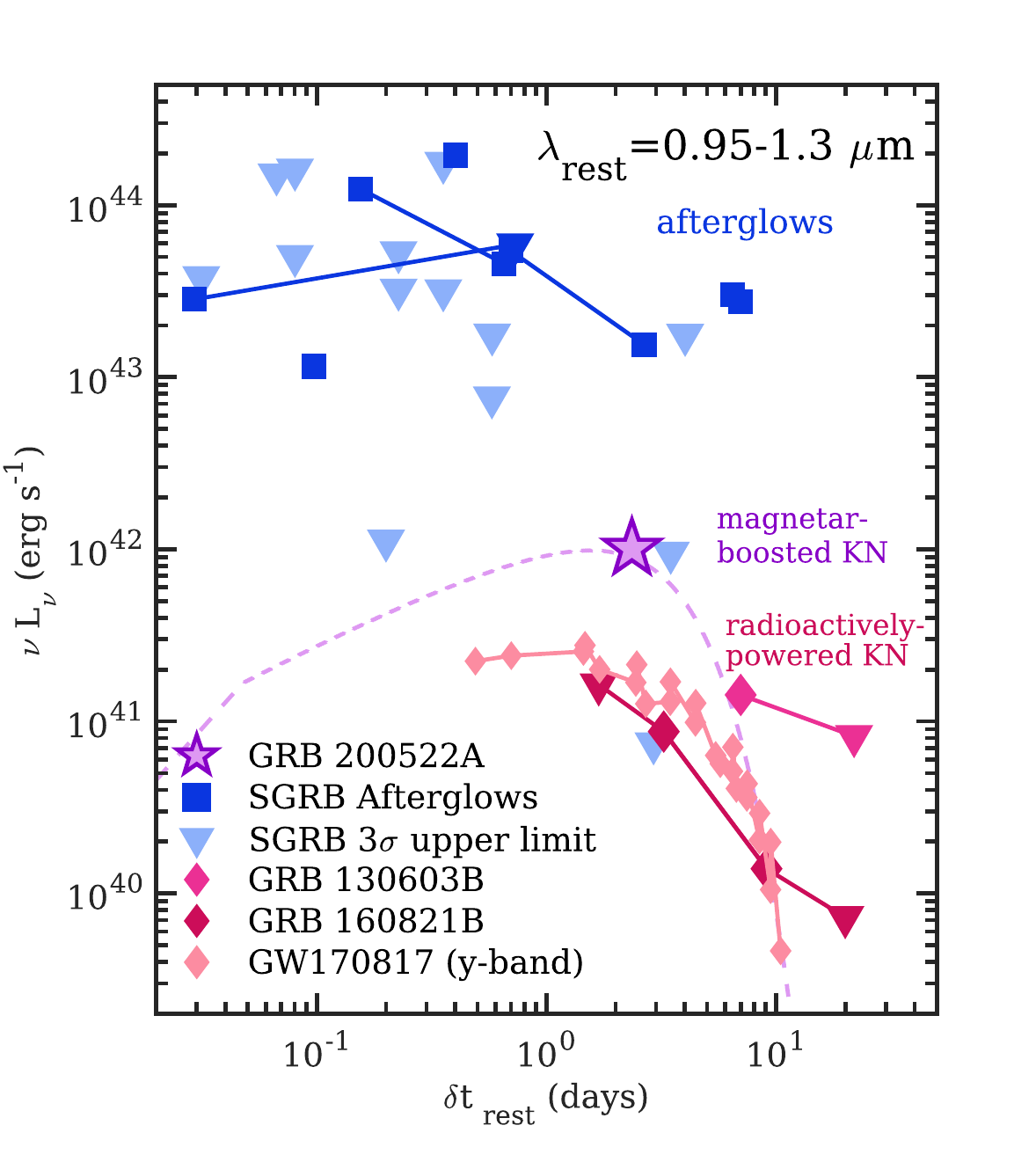}
\vspace{-0.3in}
\caption{Rest-frame $0.7-0.95\,\mu$m (left, $i$- and $z$-bands) and $0.95-1.3\,\mu$m (right; $y$ and $J$-bands) luminosity versus rest-frame time compilations. The data displayed include \grb\ (purple star), short GRB light curves including afterglow emission (blue squares), $3\sigma$ upper limits (blue triangles), and the kilonovae of GRB\,130603B \citep{bfc13,tlf+13}, GRB\,160821B \citep{ltl+19,tcb+19}, and GW170817 \citep{vgb+17}. Compared to the radioactively-powered kilonova of GW170817 the NIR counterpart of \grb\ is $\approx 8$-$17$ times more luminous. \grb\ is also significantly more luminous than other kilonova candidates in the rest-frame $i-$ and $z-$ bands and relevant times. We propose that the NIR counterpart is a kilonova with luminosity boosted energy deposition from a magnetar (``magnetar-boosted''; dashed line), or a radioactively-powered kilonova with distinct ejecta properties from previously-observed kilonovae.
\label{fig:sgrb_nirlc}}
\end{figure*}

The total observed NIR luminosity of \grb\ is $L_{\rm F125W,tot} \approx 1.7 \times 10^{42}$~erg~s$^{-1}$ and $L_{\rm F160W,tot} \approx 1.3 \times 10^{42}$~erg~s$^{-1}$ at a rest-frame time of $\delta t_{\rm rest}\approx 2.3$~days. This emission may be interpreted as originating from the forward shock of a GRB synchrotron afterglow (Scenario II in Section~\ref{sec:bb_alternative}). However, the broad-band observations require a reverse shock to explain the early radio excess, a jet break to explain the steep NIR decline, and predicts a steeper X-ray decline than the observed rate.

In this section, we further consider the implications of Scenario I, in which the radio and X-ray emission originate from a forward shock, with an excess NIR luminosity relative to this model by factors of $\approx$5--10 (Section~\ref{sec:bb_nirexcess}). We explore viable emission mechanisms that can explain the observed \grb\ F125W and F160W luminosities (corresponding to rest-frame $i$- and $y$-bands, respectively).

\subsection{An Intermediate-Luminosity NIR Counterpart}

From our modeling, we estimate that $\approx$10--30\% of the observed flux comes from the afterglow, implying a NIR {\it excess} contribution of $L_{\rm F125W,ex} \approx (9.5-12.3)\times 10^{41}$~erg~s$^{-1}$ (dropping to an upper limit of $L_{\rm F125W,ex} \lesssim 1.1 \times 10^{41}$~erg~s$^{-1}$ at $\delta t = 16.4$~days) and $L_{\rm F160W,ex} \approx (8.9-11.4)\times 10^{41}$~erg~s$^{-1}$. From the F125W and F160W observations, we also calculate a rest-frame color at $\delta t_{\rm rest}\approx 2.3$~days of $i-y=-0.08 \pm 0.21$.

To place the NIR excess emission in context with observations of other short GRBs, we collect data of all events which have observations at $\delta t_{\rm rest} \lesssim 20$~days. At $z=0.5536$, the F125W and F160W filters correspond to rest-frame wavelengths of $\lambda_{\rm rest}\approx 0.8\,\mu$m and $1.0\,\mu$m, respectively. We use observations at $\lambda_{\rm rest}=0.7-0.9\,\mu$m to compare to the F125W filter, and at $\lambda_{\rm rest}=0.95-1.3\,\mu$m to compare to F160W filter. The sources of data are the short GRB afterglow catalog \citep{fbm+15}, more recent short GRBs\,150424A \citep{jlw+18}, 150831A \citep{kkg15}, 160303A \citep{tbw+16, gbw+16}, 160410A \citep{mxk+16}, 160411A \citep{ycg16}, 170127B \citep{chm+17} and 170428A \citep{tbw+17}, and a further catalog of short GRB observations (Rastinejad et al., in prep). We also include detections of short GRBs which have been interpreted as $r$-process kilonovae, transients with thermal SEDs that result from the radioactive decay of $r$-process elements synthesized in the ejecta of a NS merger (e.g., \citealt{lp98,mmd+10,bk13}). In this vein, we include the kilonova of GRB\,130603B \citep{bfc13,tlf+13}, and the afterglow and kilonova of GRB\,160821B \citep{ltl+19,tcb+19}, both of which have data in the relevant rest-frame bands. For bursts with detections, we only include events with redshifts to enable a direct comparison between their luminosities. For upper limits, we include bursts with and without redshift information, assuming $z=0.5$ for the latter category. Finally, we include the $i$- and $y$-band light curves of the kilonova of GW170817, compiled in \citet{vgb+17} (original data from \citealt{Andreoni17,Arcavi17_2,cbv+17,cfk+17,Diaz17,Drout17,hwa+17,Pian17,scj+17,Tanvir17,Troja17,Utsumi17}). The compilation plots, along with the data of \grb, are displayed in Figure~\ref{fig:sgrb_nirlc}.

The detected NIR emission observed in \grb\ clearly lies in a unique part of parameter space. It is well below the afterglow luminosities of detected short GRBs (Figure~\ref{fig:sgrb_nirlc}), albeit with sparser sampling in the relevant bands and on the same timescales. Meanwhile, it is significantly more luminous than any known kilonova in the same rest-frame bands, which on average have $\nu L_\nu\approx 10^{41}$~erg~s$^{-1}$ at similar rest-frame times. The observed luminosities of previous short GRB-kilonovae and GW170817 match expectations for kilonovae powered by pure radioactive heating (``radioactively-powered''; Figure~\ref{fig:sgrb_nirlc}; \citealt{lp98,mmd+10,thk+14}). The NIR excess emission of \grb\ has a luminosity intermediate to detected on-axis short GRB afterglows and known kilonovae or kilonova candidates. Furthermore, we find that \grb\ is significantly bluer than GW170817, which had a color of $(i-y)=0.58 \pm 0.10$ at the same rest-frame time. Compared to GRB\,160821B, the only other short GRB-kilonova candidate with data adequate for comparison, the NIR counterpart is slightly bluer than \grb\ (with $(r-i)\approx 0.10 \pm 0.26$ and $(y-J)\approx 0.26 \pm 0.04$ at $\approx 1.7-3.3$~days; \citealt{ltl+19}), although consistent within the uncertainties.

\subsection{Radioactively-Powered Model Considerations}

\begin{figure*}
\centering
\includegraphics[width=0.4\textwidth,trim={0 0 0.3cm 0}]{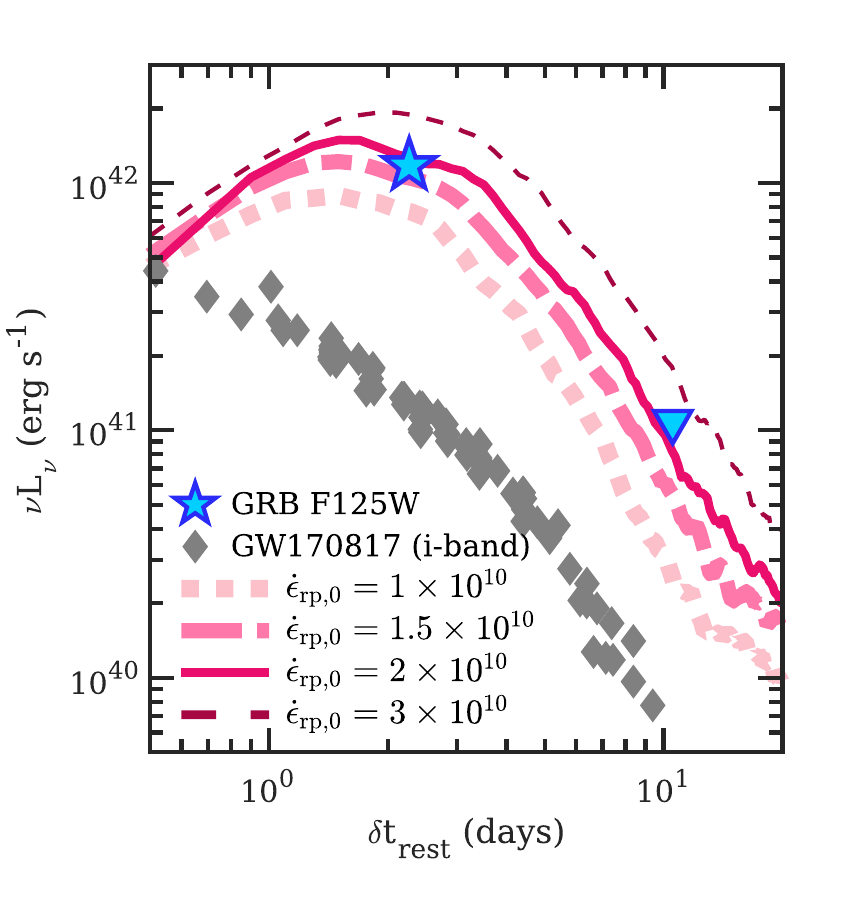}
\includegraphics[width=0.4\textwidth,trim={0.3cm 0 0 0}]{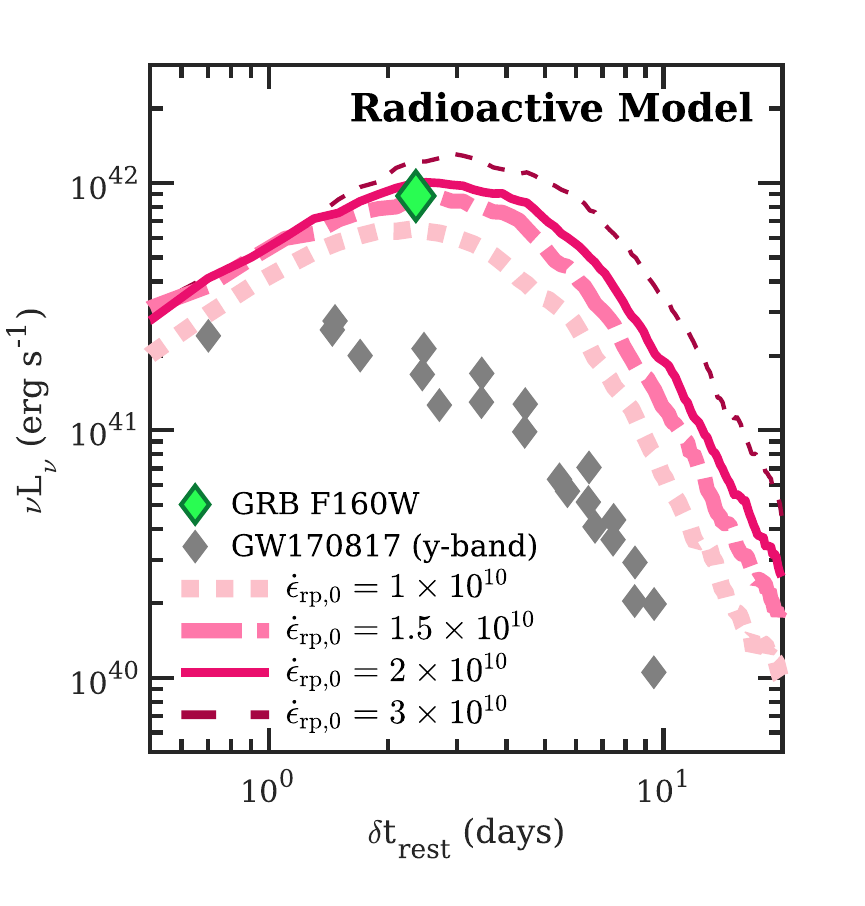}
\vspace{-0.2in}
\caption{The NIR counterpart (F125W: blue star and triangle; F160W: green diamond) of \grb\ alongside radioactive models with enhanced heating (pink lines). The four {\tt Sedona} light curve models shown assume a power-law heating rate with a range of fixed radioactive heating rate constants, $\dot{\epsilon}_{\rm rp,0}$ ($1\times10^{10}$ to $3\times10^{10}$~erg~s$^{-1}$~g$^{-1}$; pink lines), a lanthanide and actinide mass fraction of $X_{\rm lan}=10^{-3}$, ejecta mass of $M_{\rm ej}=0.1\,M_{\odot}$ and ejecta velocity of $v_{\rm ej}=0.15c$. These parameters have been chosen in attempts to match the luminosity and color of \grb; all of these models are significantly more luminous than GW170817 (gray diamonds). For these model parameters, the NIR counterpart of \grb\ requires $\dot{\epsilon}_{\rm rp,0}\gtrsim 1.5 \times 10^{10}$~erg~s$^{-1}$~g$^{-1}$, a factor of $\approx 1.9$ larger than assumed for GW170817 \citep{cbk+17,kmb+17}.
\label{fig:radioactive}}
\end{figure*}

We first explore the possibility that the luminosity and color ($i-y=-0.08 \pm 0.21$) of the NIR counterpart to \grb\ can be explained by pure \rp\ radioactive decay.
The observed NIR luminosity is ${\sim}10$ times greater than that of other known kilonovae or candidates at similar epochs (Figure~\ref{fig:sgrb_nirlc}). If attributed solely to radioactivity, this implies that the kilonova accompanying \grb\ ejected a higher mass than other kilonovae, was heated by radioactivity at a higher specific heating rate $\dot{\epsilon}_{\rm rp}$ ($[\dot{\epsilon}_{\rm rp}] = \text{erg} \, \text{s}^{-1} \, \text{g}^{-1}$)  than is commonly assumed ($\dot{\epsilon}_{\rm rp, typ}$), 
or experienced some combination of these effects, subject to the rough constraint $(M_{\rm ej}/M_{\rm ej,typ})\times(\dot{\epsilon}_{\rm rp}/\dot{\epsilon}_{\rm rp, typ}) \approx 10$.

\emph{R}-process radioactivity is generally divided into two regimes: a heavy or main \rp , and a light \rp.
The first occurs in extremely neutron-rich conditions and produces heavy elements (lanthanides and actinides) whose high opacities cause the resulting emission to peak at redder (e.g., NIR) wavelengths \citep{bk13,kbb13,th13}. In contrast, the latter, a product of relatively neutron-poor outflows, synthesizes a lighter composition with a lower opacity, leading to a transient that generally peaks at bluer (optical) wavelengths.
Though GW170817 showed evidence of both a light and a main \rp\ 
\citep{vgb+17,kmb+17,met19}, the bluer color of \grb\ suggests its emission is dominated by a light \rp, low-opacity component.
This is not unexpected for kilonovae viewed from the polar direction \citep{skk+16,wsn+14,mb14,prc+14,bkw+16,Kilpatrick17}, or whose central remnants are long-lived NSs.
In the latter case, neutrino irradiation of the accretion disk by the central NS will raise the electron fraction (\ye; the number of electrons per baryon) of outflowing disk material, inducing a light \rp\ \citep{mf14,kfm15,lfr+17}. Magnetar winds from the NS surface can provide additional high--\ye, low-opacity material \citep{mtq18}.

The apparent low opacity complicates the question of enhanced \rp\ heating for \grb. There is some variability in predictions of \rp\ heating rates, due to the uncertain physics of the neutron-rich nuclei involved and the diverse astrophysical conditions that may characterize an \rp\ event \citep[see, e.g.][]{bkw+16}. However, these uncertainties are greatest for the heaviest nuclei, while the relatively blue color of the NIR counterpart to \grb\ 
suggests a \rp\ that failed to fuse many elements with $A \gtrsim 130$, and a light \rp.
(The higher temperatures associated with higher specific heating rates could in theory push the thermal SED blueward, reproducing the blue colors without the requirement of low opacity.
However, we found that absent an extreme choice of heating rate, this effect was too small to overcome the reddening from from high-opacity lanthanides and actinides if these are present at mass fractions greater than $X_{\rm lan}\sim10^{-3}$.)
If the NIR counterpart is to be explained by pure radioactive decay, the observed color seems to require a weak (low-lanthanide) \emph{r}-process.

As a test case, we consider an outflow with ejecta mass $M_{\rm ej} = 0.1M_\odot$, average ejecta velocity $v_{\rm ej} = 0.15 c$, and a combined lanthanide and actinide mass fraction of $X_{\rm lan} = 10^{-3}$. 
This could be considered a pure-radioactive energy analog to the magnetar-boosted model (Section~\ref{sec:magnetar_kn}).
Such a scenario might arise if a NS central remnant survived long enough to neutrino-irradiate its accretion disk and drive the material to a high $Y_{\rm e}$ (e.g., \citealt{lfr+17}), but not long enough to impart its spin-down energy to the ejecta
(however, see also  \citealt{mrd+19}, who
suggest that a central NS may not be necessary for a high-$Y_{\rm e}$ disk outflow).
We simulate the resulting emission using the radiation transport code \texttt{Sedona} \citep{ktn06}, parametrizing the \rp\ heating rate with a power law, 
\begin{equation}
\dote = \doteo(t/\text{day})^{-1.3}.
\label{eq:plaw_heat}
\end{equation}
The power-law index $\alpha = 1.3$ is a standard analytic approximation for \rp\ heating.
It is expected from Fermi's theory of $\beta$-decay  \citep{hsp17,kb19},
and has been shown to be consistent with the results detailed numerical models of the \rp\ \citep[e.g.,][]{mmd+10, kraw12}. Typical values for \doteo\ are ${\sim}10^{10}$ erg s$^{-1}$ g$^{-1}$. Here, we consider a range of models, from $\dot{\epsilon}_{\rm rp,0}=(1-3) \times 10^{10}$~erg~s$^{-1}$~g$^{-1}$ (Figure~\ref{fig:radioactive}).

While not all of the energy released by the \rp\ is actually available to power the kilonova's electromagnetic emission, due to inefficient thermalization of radioactive energy \citep{bkw+16},
thermalization is efficient at early times and for more massive and/or slower-moving ejecta.
We therefore absorb the effects of thermalization into Eq.~\ref{eq:plaw_heat} and assume in our radiation transport calculation that all emitted energy is efficiently absorbed.

Our radioactively-powered model is able to reproduce both the color and the observed $i$- and $y$-band luminosities of the NIR counterpart of \grb\ only for $\doteo \gtrsim 1.5 \times 10^{10}$ erg s$^{-1}$ g$^{-1}$ (Figure~\ref{fig:radioactive}).
This is a factor of ${\gtrsim}1.5$ higher than what has typically been assumed.
For example, the kilonova models of \citealt{kmb+17,cbk+17} to explain GW170817
had an effective heating rate (including thermalization) approximately equal to $8 \times 10^{9} \; (t/\text{day})^{-1.3}$ erg s$^{-1}$ g$^{-1}$ for $0.1 \leq t/\text{day} \leq 5$, lower than the model that can explain \grb\ by a factor of $\sim 1.9$.

Assuming that $\beta$-decays supply most of the radioactivity, and that the difference between emitted and thermalized radioactive energy is due only to neutrinos, which carry away ${\sim}1/3$ of the energy of a typical $\beta$-decay, our results suggest a true \rp\ heating rate of $\dote \approx 2.3\times 10^{10} \, (t/\text{day})^{-1.3}$ erg s$^{-1}$ g$^{-1}$. 
In summary, if the NIR emission of \grb\ is produced by a radioactively-powered kilonova, the properties of this ejecta (e.g., mass, heating, and/or composition) must be different than those inferred for GW170817. Detailed models exploring these properties, coupled to more-detailed heating prescriptions, are required to fully understand the NIR counterpart of \grb\ in the context of radioactive models, as well as implications for other kilonovae.

\subsection{Magnetar-boosted Kilonova Model}
\label{sec:magnetar_kn}
As described in the previous section, the NIR emission and color of \grb\ are difficult to explain by a radioactive heating alone, under standard assumptions about ejected mass and the specific heating from $r$-process decay. However, it is possible that deposition of energy from a NS remnant created as a result of the merger can boost the optical and NIR luminosity of the kilonova by up to a factor of $\approx 100$ (``magnetar-boosted'' kilonova; \citealt{yzg13,mp14}; see also \citealt{kin16,mik+18} for general ``engine-powered" models). Indeed, a small fraction of BNS mergers are expected to produce a supramassive NS remnant that is indefinitely stable to collapse (e.g.~\citealt{mm19}). The remnant may acquire large magnetic fields during the merger process and is necessarily spinning near break-up (e.g.~\citealt{sch+13,kks+18}), resulting in a rapidly-spinning ``magnetar'', which provides a reservoir of energy via spin-down that is not available in the scenario of a prompt collapse to a black hole. Since the kilonova ejecta mass is expected to be of order $M_{\rm ej}\approx$0.01--0.1$\,M_{\odot}$ \citep{met19}, in this scenario, the rotational energy is deposited behind the ejecta into an expanding nebula with a non-thermal component in the X-ray band and a thermal component peaking at optical and NIR wavelengths.

\begin{figure}
\centering
\includegraphics[width=0.5\textwidth]{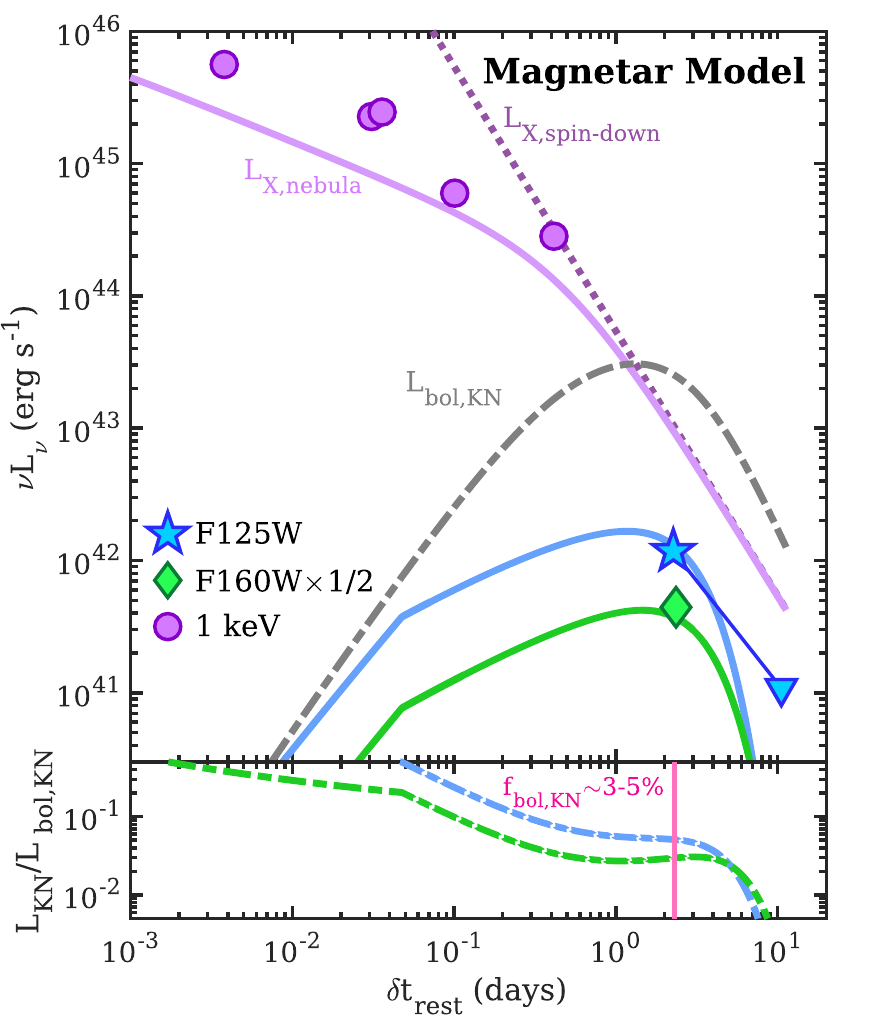}
\vspace{-0.2in}
\caption{The X-ray afterglow light curve and NIR excess of \grb\ (purple, blue and green points); triangles represent $3\sigma$ upper limits. Also shown are the predicted emission contributions of a magnetar model with $B=2.5 \times 10^{15}$~G, $P=0.7$~ms and $\kappa=1$~cm$^2$~g$^{-1}$ \citep{mp14,met19}. A fraction of the spin-down luminosity (dotted line) powers the non-thermal nebular X-rays (purple solid line), the latter of which is predicted to be sub-dominant compared to the forward shock afterglow. The nebular emission is also thermalized into an optical/NIR ``magnetar-boosted'' kilonova with a peak bolometric luminosity of $\approx 3 \times 10^{43}$~erg~s$^{-1}$ (dot-dashed gray curve). The contribution in the F125W and F160W bands (rest-frame $i$- and $y$-bands) are shown as solid lines. The bottom panel illustrates the fraction of luminosity in two {\it HST} filters contributing to the bolometric kilonova luminosity, $\approx 3-5\%$ at the time of the {\it HST} detections.
\label{fig:magnetar}}
\end{figure}

We investigate the feasibility that the NIR excess emission of \grb\ can be explained by a magnetar-boosted kilonova. Using the formalism presented in \citet{met19} (accounting for corrections to the effective engine luminosity from \citealt{mp14}), we fix the opacity to $\kappa=1$~cm$^2$~g~$^{-1}$ (corresponding to an electron fraction, $Y_e \approx 0.4$, in the ``blue'' regime), as was found to explain the early blue emission of GW170817 \citep{tkg+20}. We employ light curve models with magnetic field strengths of $B=(2.5-3) \times 10^{15}$~G, initial spin period $P_0 = 0.7$ ms (corresponding roughly to the break-up rate), and a total ejecta mass of $M_{\rm ej}=0.1\,M_{\odot}$ (similar to the disk wind ejecta in the case of a long-lived neutron star; e.g.~\citealt{mf14}). The spin-down luminosity ($L_{\rm X,sd}\propto t^{-2}$) provides an energy reservoir, which powers the expanding nebula, and which is thermalized at optical and NIR wavelengths. The nebula is not expected to be transparent to X-rays until the ejecta are ionized (on $\gtrsim 1$ to few-day timescales). A comparison of our model to the X-ray observations of \grb\ demonstrates that the predicted nebular X-ray emission is a factor of $\approx 2$ below the observed values (Figure~\ref{fig:magnetar}), although does have a similarly shallow decline rate at $\lesssim 0.4$~days of $L_{\rm X,neb} \propto t^{-0.6}$. Thus, the observed X-ray emission of \grb\ is likely to be dominated by the FS afterglow emission in this model. We note that the NIR photons from the nebula may provide an additional source of cooling for X-ray synchrotron-emitting electrons at the FS. However, for the high Lorentz factor of the FS at the time of the X-ray observations ($\delta t\lesssim3.5$~days; $\Gamma\gtrsim6$), this effect is negligible even for the high NIR photon density inferred here \citep{ls19}.

We find that the magnetar model matches the colors and luminosity of the NIR excess emission (Figure~\ref{fig:magnetar}). For these parameters, the peak of the kilonova SED is significantly bluer than our observing bands: at $\delta t_{\rm rest} \approx 2.3$~days, the effective temperature is $T_{\rm eff}\approx6430-6960$~K, corresponding to $\lambda_{\rm pk}\approx0.42-0.45\,\mu$m. Thus, our {\it HST} observations only account for $\approx 3-5\%$ of the predicted bolometric kilonova luminosity at that time (Figure~\ref{fig:magnetar}). 

\subsection{Comparison to Short GRBs and GW170817}

\begin{figure*}
\centering
\includegraphics[width=0.8\textwidth]{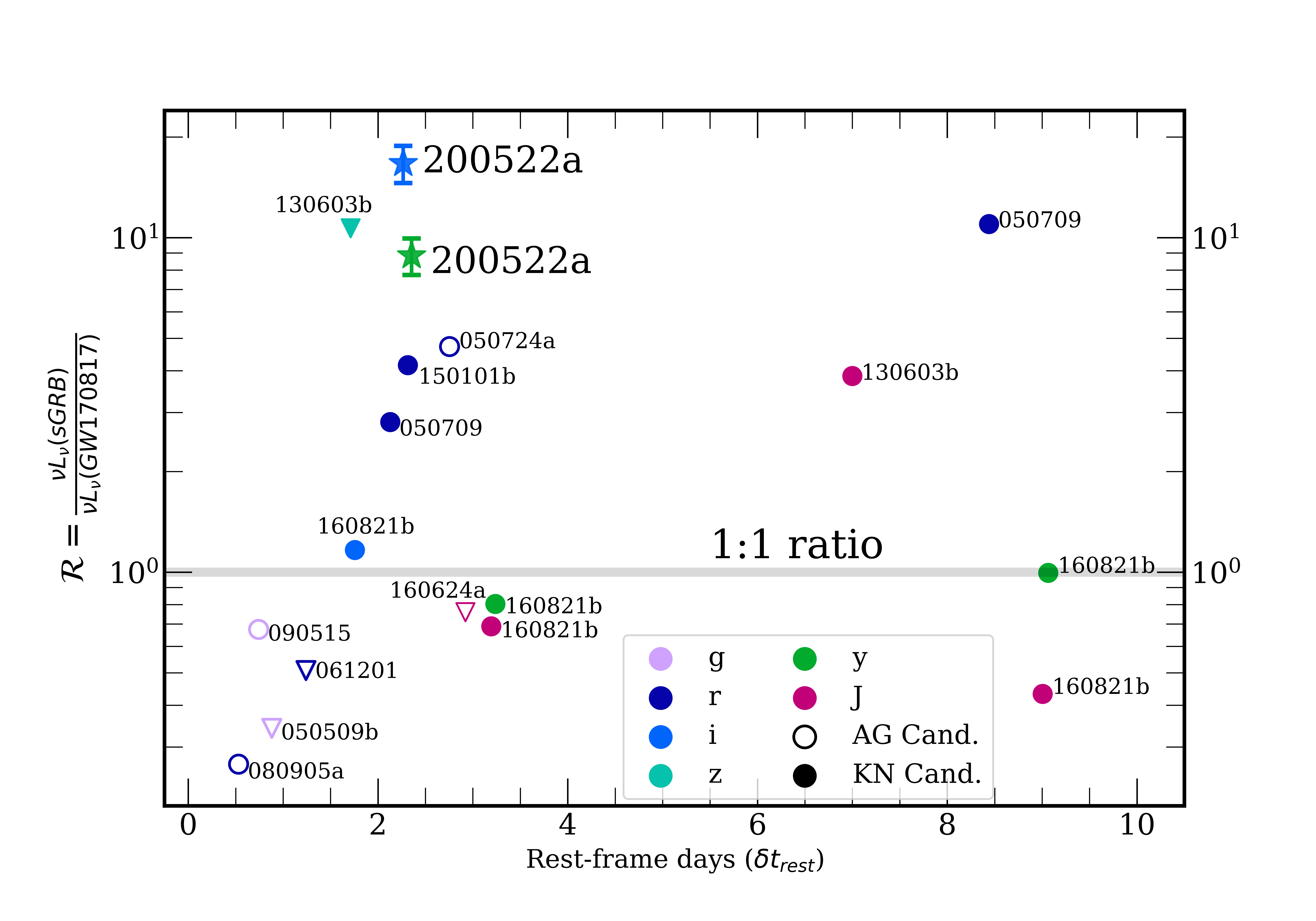}
\vspace{-0.2in}
\caption{Ratios of relevant SGRB observations to the lightcurve of GW170817 in restframe time, restframe band and luminosity, $\nu L_{\nu}$. Observations are color coded by rest-frame band. Open circles denote the ratios of SGRBs afterglows that have not been claimed as kilonovae. Closed circles mark the ratios of SGRB kilonovae detections. Triangles show the ratios of SGRB kilonovae upper limits. The gray horizontal line denotes a lightcurve equal to GW170817's kilonova ($\mathcal{R}=1$) against which each short GRB observation can be independently compared. Most previous claimed kilonova fall within a factor of 10 of GW170817 but show diversity in color and luminosity. Early {\it HST} detections of \grb, marked as stars, appear 16.7 and 8.8 times as luminous as GW170817 in rest-frame i- and y-bands respectively.
\label{fig:kn_ratio}}
\end{figure*}

In the context of interpreting the NIR excess emission of \grb\ as a kilonova, we are thus motivated to directly compare the NIR emission to that of GW170817, and to the landscape of short GRBs with optical or NIR emission (or limits) within $\approx 10$~times the luminosity of GW170817 across all observed bands (Figure~\ref{fig:kn_ratio}).

Our comparison sample of relevant short GRB consists of GRBs\, 050709 \citep{jhl+16}, 130603B \citep{tlf+13,bfc13}, 1501010B \citep{fmc+16}, and 160821B \citep {ltl+19,tcb+19}. For GRB\,160821B we include only optical detections at $1.75 \lesssim \delta t_{\rm rest} \lesssim 5$ days and NIR detections $\delta t_{\rm rest} \gtrsim 1.5$ days, where the kilonova emission was found to dominate the afterglow \citep {ltl+19,tcb+19}. We also include highly-constraining afterglow upper limits (e.g., GRBs\,050509B, \citealt{csb+05}; 061201, \citealt{fbm+15}; 160624A) and low-luminosity short GRB afterglows that do not have existing kilonova interpretations (GRBs\,050724A, \citealt{bpc+05}; 080905A, \citealt{rwl+10}; 090515, \citealt{rot+10}). Each short GRB has a clear, well-measured redshift that allows us to calculate accurate luminosities. For each of the bursts, we select the most relevant or constraining observations available in the observed $grizyJ$-bands.

For GW170817, we make use of the available multi-band light curves compiled in \citet{vgb+17}, performing a linear interpolation in 1-hour time bins, transforming them to rest-frame luminosities and times. Similarly, we transform each of the short GRB observations to their rest-frame wavelengths, luminosities and times. For each short GRB observation, we compute the ratio of luminosities, $\mathcal{R}=\nu L_{\nu}$(SGRB)/$\nu L_{\nu}$(GW170817), at the relevant rest-frame time. We show the ratio $\mathcal{R}$ versus rest-frame time. The gray horizontal line represents a 1:1 ratio ($\mathcal{R}=1$) against which each short GRB observation can be independently compared.

It is clear that the NIR excess observed in \grb\ is significantly more luminous than candidate kilonovae and GW170817 (Figure~\ref{fig:kn_ratio}). The color evolution from blue to redder bands over time as expected for kilonovae is overall apparent. The NIR counterpart of \grb\ at $\delta t_{\rm rest}\approx 2.3$~days is significantly brighter than GW170817 with $\mathcal{R}\approx 16.8$ and 8.8 in the $i$- and $y$-bands, respectively. These ratios are also significantly higher than $\mathcal{R} \approx 4$ for the candidate kilonovae of GRBs\,130603B and 150101B \citep{bfc13,tlf+13,trp+18}. \grb\ is $\approx 9.3$ and $\approx 13.2$ times more luminous than GRB\,160821B, the only short GRB-kilonova candidate for which data exist at similar rest-frame times and bands. Overall, Figure~\ref{fig:kn_ratio} highlights the diversity of late-time excess emission in short GRBs in terms of luminosities and colors (see also: \citealt{glt+18,acd+19,rsm+20}). It also highlights the effectiveness of searches traditionally fine-tuned for afterglows in reaching the depths required to detect nearby ($z \lesssim 0.3$) kilonovae similar to the luminosities of GW170817.

\section{Discussion}
\label{sec:discussion}

\subsection{The Host Galaxy of \grb\ in Context}
First, we examine the host of \grb\ in the context of the short GRB population and field galaxies. \grb\ is located at a small projected physical offset of $\approx 1$~kpc, or $\approx 0.24r_e$ from the center of its host galaxy, closer than 90\% of short GRBs \citep{fb13}. The location of \grb\ is also indicative of a strong correlation with its host stellar mass distribution, residing at the 95\% level in terms of its host rest-frame optical light. However, the low afterglow-inferred circumburst density of $\approx 10^{-3}-10^{-2}$~cm$^{-3}$ is somewhat surprising given its placement in its host galaxy (modulo projection effects); indeed the inferred value is in line with the typical expected densities of short GRBs, the majority of which occur at significantly larger offsets. The host galaxy also exhibits an asymmetric morphology with a bulge and a disturbed disk, potentially indicative of a recent merger or fly-by encounter.

Compared to the host galaxies of other short GRBs, the host of \grb\ comprises a fairly young, low-mass stellar population, falling in the lower $38\%$ and $25\%$ of all short GRB host stellar masses and ages that have been derived in a similar manner \citep{nfd+20}. Compared to the galaxy luminosity function at this redshift, the host galaxy has a luminosity $\approx$ 0.5$L^*$ \citep{Willmer2006}, on the low end for short GRB hosts. Approximately $70\%$ of short GRB host galaxies have evidence of ongoing star formation \citep{fbc+13}, with a median $\text{SFR} \approx 1\,M_\odot$~yr$^{-1}$ \citep{ber14}; in comparison, the host of \grb\ is more strongly star-forming than most short GRB hosts, with $\text{SFR} \approx 2.1-4.8\,M_\odot$~yr$^{-1}$. However, compared to field galaxies of similar stellar mass at $0.5 < z < 1$, the host is consistent with or just below the main locus of star-forming galaxies on the main sequence, depending on where in the range the true SFR is \citep{Whitaker2014,Fang2018}. This means that given its stellar mass, the host of \grb\ is forming stars comparable or at a slightly lower rate than contemporary field galaxies.

\subsection{Precursor Emission, Radio Afterglows and Reverse Shocks in Short GRBs}
\begin{deluxetable}{lcccc}
\linespread{1.2}
\tablecolumns{5}
\tablewidth{0pc}
\tablecaption{Radio GRB Afterglows
\label{tab:radio}}
\tablehead {
\colhead {GRB}                &
\colhead{$\delta$t}                 &
\colhead{Frequency}             &
\colhead {Afterglow} &
\colhead{Ref} \\
\colhead {}           &
\colhead {(days)}      &
\colhead {(GHz)}      &
\colhead {($\mu$Jy)}    & 
\colhead{}
}
\startdata
\hline
GRB\,050724A & 0.57 & 8.46 & 173$\pm$30 & 1 \\
 & 1.69 & 8.46 & 465$\pm$29  \\ 
\hline
GRB\,051221A & 0.91 & 8.46 & 155$\pm$30 & 2 \\
 & 1.94 & 8.46 & $\lesssim$72  \\
 & 3.75 & 8.46 & $\lesssim$96  \\
 & 6.88 & 8.46 & $\lesssim$84  \\
 & 23.93 & 8.46 & $\lesssim$48  \\ 
\hline
GRB\,130603B & 0.37 & 6.7 & 119$\pm$9.1 & 3 \\
 & 1.43 & 6.7 & 65$\pm$15.2  \\
 & 4.32 & 6.7 & $\lesssim$26  \\ 
\hline
GRB\,140903A & 0.404 & 6.0 & 110$\pm$9.5 & 4 \\ 
 & 2.45 & 6.0 & 187$\pm$8.7  \\ 
 & 4.7 & 6.0 & 127.9$\pm$15.1  \\ 
 & 9.24 & 6.0 & 81.9$\pm$14.7  \\ 
 & 18.24 & 6.0 & $\lesssim$120 \\ 
\hline
GRB\,141212A & 0.45 & 6.0 & $\lesssim$25.2 & 5 \\
 & 3.76 & 6.0 & 27.0$\pm$8.1  \\
 & 7.72 & 6.0 & 21.3$\pm$6.4  \\ 
\hline
GRB\,150424A & 0.77 & 9.8 & 32.8$\pm$8.9 & This work\\
 & 4.69 & 9.8 & $\lesssim$18.6  \\
 & 7.90 & 9.8 & $\lesssim$12.9  \\
 & 6.29$^a$ & 9.8 & $\lesssim$11.4  \\ 
\hline
GRB\,160821B & 0.17 & 5.0 &  40.1$\pm$8.9  & This work\\
 & 1.12 & 5.0 &  $\lesssim$16.5  \\
 & 10.06 & 9.8 & $16.0\pm4.0$ & 6 \\
 & 17.09 & 9.8 & $<33.0$ & 6 \\ 
\hline
GRB\,200522A & 0.23 & 6.05 & 33.4$\pm$8.2 & This work \\
 & 2.19 & 6.05 & 27.1$\pm$7.2   \\
 & 2.19 & 9.77 & $\lesssim$23.7   \\
 & 6.15 & 6.05 & $\lesssim$18.6   \\
& 11.15 & 6.05 & $\lesssim$14.1  \\
\enddata
\tablecomments{Uncertainties correspond to 1$\sigma$ confidence and upper limits correspond to 3$\sigma$ \\
$^a$ Combination of 9.8 GHz observations at 4.69 days and 7.90 days \\
{\bf References:} (1) \citealt{bpc+05}, (2) \citealt{sbk+06}, (3) \citealt{fbm+14}, (4) \citealt{fbm+15}, (5) This work, (6) \citealt{ltl+19} }
\end{deluxetable}

\begin{figure}
\centering
\includegraphics[width=0.495\textwidth]{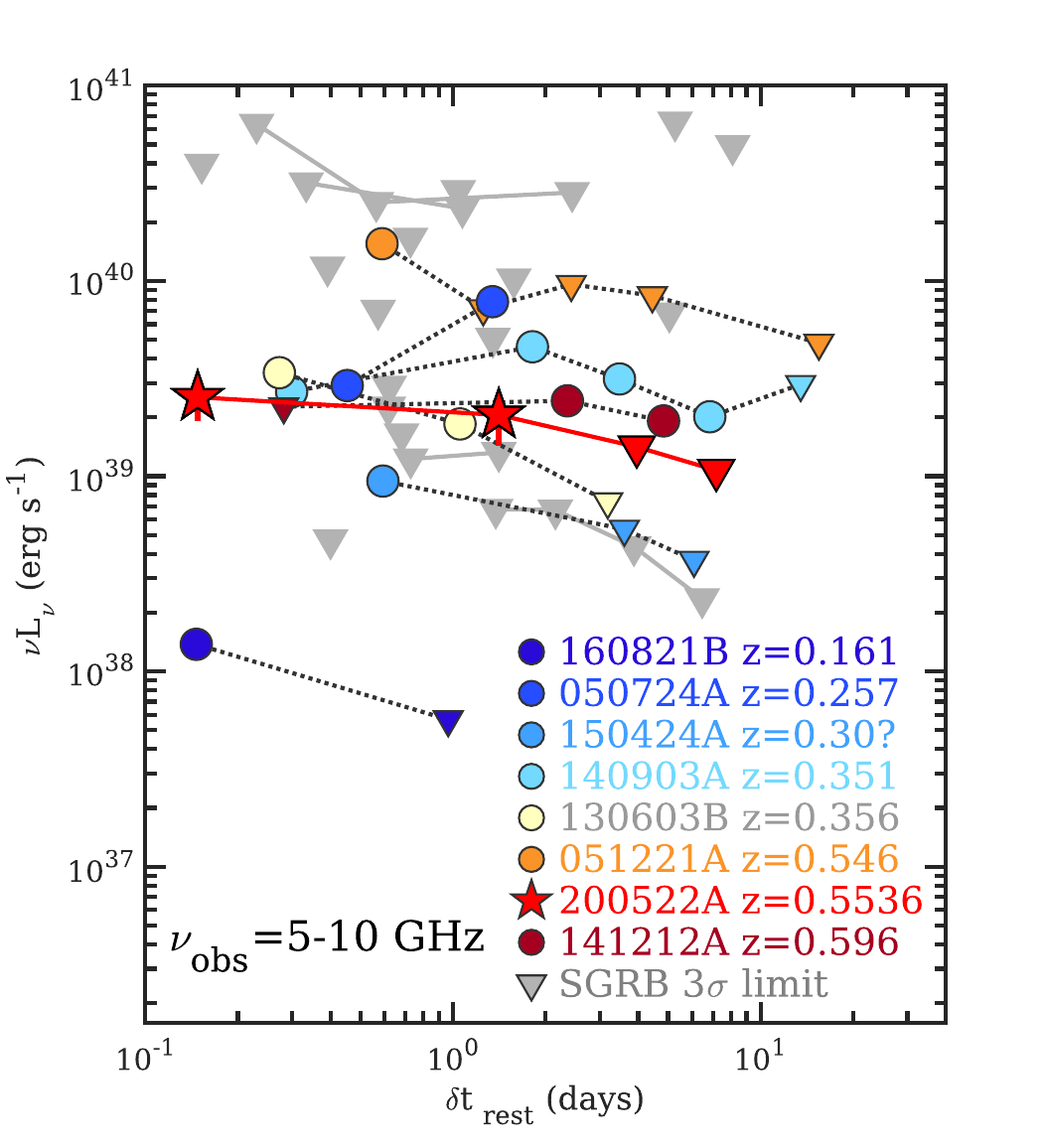}
\vspace{-0.3in}
\caption{Radio luminosity ($\nu L_\nu$) of the afterglow of \grb\ (star points) versus rest-frame time. Also shown are the seven additional short GRB afterglow detections to date with GHz observations (circles). Lines connect data points for the same burst and triangles denote $3\sigma$ upper limits. Bursts are color-ordered by their host galaxy redshift from low-redshift (blue) to higher redshifts (red). 
\label{fig:radlc}}
\end{figure}

We now place the broad-band properties of \grb\ and its host galaxy in the context of the short GRB population. The possible presence of $\gamma$-ray precursor emission on timescales of $<1$~second of the main pulse of \grb\ is intriguing, given that only $\approx 10\%$ of {\it Swift}/BAT short GRBs have been found to have such emission \citep{trg10}. Furthermore, most short GRBs with precursor emission had significantly longer quiescence timescales of tens of seconds between the precursor and the GRB; only one other event, GRB\,090510, had a detected precursor within $1$~second. The physical origin of pre-cursor emission is unknown. Theroetical models include the excitement of tidal resonances between the component neutron stars during the merger \citep{trh+12,sk20}, or accretion onto a magnetar central engine (e.g., \citealt{bcg+13}).

Turning to the afterglow emission, the radio afterglow of \grb\ represents the eighth radio afterglow detection for a short GRB out of a total of $>70$~events observed. The lack of radio detections has been attributed to the relatively lower energy scales and circumburst densities \citep{fbm+15} compared to their long GRB counterparts \citep{pk02,yhs+03,cfh+10,cfh+11,lbt+14,lbm+15}. Using the redshift of \grb, the radio afterglow luminosity is $\nu L_{\nu}=(2.5 \pm 0.6) \times 10^{39}$~erg~s$^{-1}$ at $\delta t_{\rm rest}=0.15$~days, and the radio counterpart was detected through $\delta t_{\rm rest}=1.4$~days. To compare the luminosity and behavior to those of other radio afterglows, we collect available radio afterglow data taken at 5-10~GHz frequencies for short GRBs with redshifts. For the radio afterglow detections, we gather data for GRBs\,050724A \citep{bpc+05}, 051221A \citep{sbk+06}, 130603B \citep{fbm+14}, 140903A \citep{tsc+16}, 141212A \citep{fbm+15}, and 160821B (9.8~GHz; \citealt{ltl+19}). In addition, we reduce and analyze 9.8~GHz observations for GRBs\,150424A and $5.0$~GHz data for 160821B (Program 15A-235, PI:~Berger; \citealt{gcn17804,gcn19854}) and present their fluxes and upper limits here. Finally, we include upper limits for 18 short GRBs with redshifts from \citet{fbb+17}. The total sample of short GRB radio afterglows with redshifts comprises 27 events, and their radio luminosity light curves are shown in Figure~\ref{fig:radlc} and listed in Table~\ref{tab:radio}.

For the detections, the redshifts span $z=0.16-0.596$, tracing the low-redshift end of the distribution of short GRBs \citep{pfn+20}, which can be attributed to observational selection effects. While \grb\ is among the most distant radio afterglow detections, we find that its luminosity is unexceptional, and squarely in the range of those traced by short GRBs, which have $\nu L_{\nu}\approx 10^{39}-10^{40}$~erg~s$^{-1}$. The one exception is GRB\,160821B, whose radio afterglow was an order of magnitude less luminous than the other GRBs; together with its multi-wavelength data, that event was interpreted as a slightly off-axis structured jet \citep{tcb+19} or the result of a narrow jet with a reverse shock \citep{ltl+19}. Finally for context, the peak radio luminosity of the off-axis afterglow of GW170817 was $\approx 8 \times 10^{35}$~erg~s$^{-1}$ at $\delta t \approx 160$~days \citep{amb+18,dkm+18,max+18}, well below those of on-axis short GRB afterglows. We also note that the X-ray afterglow of \grb\ falls just below the median luminosity for XRT afterglows. Overall, the radio and X-ray emission of \grb\ seem to exhibit similar behavior to those of on-axis short GRB afterglows. 

One of the ways to explain the multi-wavelength radio to X-ray light curves of \grb\ is through the standard synchrotron forward shock model, together with a reverse shock and a jet break. Reverse shocks are expected in weakly magnetized, baryonic ejecta, and provide a means to infer the jet initial Lorentz factor ($\Gamma_0$) and the relative magnetization ($R_{\rm B}$) of the ejecta \citep{sp99,hk13}. As the RS peak frequency is suppressed by a factor of $\Gamma_0^2$ relative to the FS, the RS is expected to be more easily detectable at radio frequencies \citep{ks00,kmk+15}. This has been borne out by observations of long-duration GRBs with the VLA, revealing a wide diversity in initial Lorentz factors ($\Gamma_0\approx100$--300) and magnetization properties ($R_{\rm B}\approx0.5$--10; \citealt{lbz+13,pcc+14,lab+16,alb+17,lab+18,lbm+18,lag+19,lves+19}).

Similarly, reverse shocks have been used to explain the early-time radio and optical excesses at $\lesssim 1$~day in three short GRBs to date. GRBs\,051221A \citep{sbk+06} and 160821B \citep{ltl+19,tcb+19} each exhibited radio excess emission relative to the forward shock model, followed by subsequent fading, while for the more recent GRB\,180418A, a reverse shock was invoked to explain an excess of optical emission at early times \citep{bdw+19}. For \grb, the reverse shock interpretation is driven by the early radio emission. 

We interpret the steep NIR decline as post jet-break behavior with a jet break at $t_{\rm jet} \approx 3.4$~days, leading to a relatively wide opening angle of $\approx 14^{\circ}$. Two other short GRBs with RS signatures, GRB\,051221A and 160821B, also had temporal steepenings in their light curves interpreted as jet breaks, leading to opening angles of $\approx 7^{\circ}$ and $\approx 2-8^{\circ}$ respectively \citep{sbk+06,ltl+19,tcb+19}. If this interpretation for \grb\ is correct, this would be the widest jet measurement that exists for a short GRB, as short GRBs with measured jets have inferred $\approx 2-8^{\circ}$ (median of $6 \pm 1^{\circ}$; \citealt{fbm+15}). In addition, only a few events have comparable lower limits indicative of wider jets, including GRB\,050709, 050724A, and 120804A with $\gtrsim 13-25^{\circ}$ \citep{gbp+06,whj+06,bzl+13}.

\subsection{An Observational Test of the Magnetar Model and Implications for Future Detectability}

Another way to understand the multi-frequency light curves and SEDs of \grb\ is by interpreting the NIR emission as a luminous kilonova. While the NIR detections of \grb\ are fainter than any on-axis afterglow detected to date at these epochs, they are a factor of $\approx 8-17$ times the luminosity of GW170817, and more luminous than any known kilonova or kilonova candidate across all observing bands (Figure~\ref{fig:kn_ratio}). Deep observations of short GRBs on the same timescales have ruled out emission with similar luminosities to the NIR counterpart to \grb\ for only two other events (Figure~\ref{fig:sgrb_nirlc}). 
We find that such a luminous NIR counterpart could be driven by heating from the spin down of a nascent magnetar or through a radioactively-powered model with enhanced specific heating rates, (a factor of $\gtrsim 2$~larger than that assumed for GW170817), a low-lanthanide composition, and a fairly high ejecta mass.

If the progenitor of \grb\ indeed produced a magnetar that is stable to collapse, synchrotron radio emission resulting from the interaction between the expanding ejecta and the surrounding medium is predicted on a few $\approx$~year timescales \citep{mb14,hp15,lgz20}. Future radio observations offer a concrete way to test the magnetar-boosted kilonova interpretation for \grb. Previous surveys searching for late-time radio emission in short GRBs have resulted in non-detections \citep{fmb+16,hhp+16,knm+19,smf+20} and an inference on the fraction of short GRBs which produce stable magnetars of $\lesssim 50\%$ \citep{smf+20}.

We use the light curve modeling described in \citet{smf+20} for an energy deposition of $10^{53}$~erg representing the maximum energy extractable from a stable remnant, as is expected to explain the magnetar-boosted kilonova interpretation for \grb. We fix the median parameters from the forward shock model ($\epsilon_B=0.01$). For a fixed ejecta mass of $M_{\rm ej} = 0.03 M_\odot$ ($0.1 M_\odot$), we find that the 6~GHz radio emission will peak at $\delta t \approx 1.5$~years ($\approx 9.9$~years) after the burst with a flux density of $F_{\nu} \approx 180\,\mu$Jy ($\approx 25.3\,\mu$Jy). Due to the rising light curve, with a peak corresponding to the deceleration timescale (e.g. \citealt{np11}), the radio emission from \grb\ will be detectable with the VLA at much earlier times than the peak, reaching $F_{\nu}\approx 20\,\mu$Jy at $\delta t\approx 0.3-6.0$~years depending on the ejecta mass. The detection of radio emission from \grb\ would be a ``smoking gun'' of this scenario and the first possible evidence of a stable magnetar created as a result of a short GRB.

If the NIR counterpart of \grb\ is relatively isotropic, the larger luminosity compared to GW170817 has implications for detectability following gravitational wave (GW) events. Most optical searches following GW events reach depths of $\approx 21$--$22$~mag (e.g., \citealt{hgc+19,lpf+19,kaa+20}). Assuming that the required depth of a search is $\approx 10$ times below peak brightness for robust counterpart detection, kilonovae of comparable brightness to GW170817 are detectable to $\approx 60$--$100$~Mpc. In comparison, high-luminosity ($\approx 10^{42}$~erg~s$^{-1}$) counterparts like that of \grb\ will be detectable by current GW counterpart search efforts to $\approx 160$--$250$~Mpc, well-matched to the expected GW network reach of BNS mergers in the O4 observing run \citep{GWProspects}, and to $\approx 600$~Mpc with the Vera Rubin Observatory (VRO; \citealt{LSST}). This is well beyond the expected GW detectability of BNS mergers during the O5 observing run. However, only a small fraction of BNS mergers are expected to produce stable magnetars (\citealt{mm19}; see also: \citealt{smf+20} for short GRBs), and thus the expected fraction of high-luminosity counterparts may also be low, if indeed the NIR couterpart of \grb\ was a result of a stable magnetar.

However, alternative and relatively unexplored explanations which are independent of a stable remnant remain, including variations to the radioactive heating rate, or speculative sources of ejecta heating such as disk winds powered by fall-back accretion (which could vary depending on the amount of fall-back; e.g., \citealt{kit15,met19}. Moreover, any modifications to radioactive heating prescriptions would necessarily need to be investigated in the context of all detected kilonovae. Future broad-band campaigns following low-$z$ short GRBs will help elucidate the nature and prevalence of the unusual emission of \grb\ and in turn the implications on detectability following GW events.

\section{Conclusions \& Future Outlook}
\label{sec:conc}
We have presented multi-wavelength observations of the counterpart of \grb\ and its host galaxy using {\it Swift}/XRT, VLA, {\it HST}, Keck, LCOGT, and archival data. We present modeling results of the afterglow and host galaxy, and propose scenarios to explain the unusual broad-band emission of \grb.

Against the backdrop of 15 years of {\it Swift} short GRB afterglow discoveries, \grb\ represents a remarkable example of the diversity of observed behavior in short GRBs. The detected luminosity of the NIR (rest-frame optical) emission on timescales of $\approx$few days, during which extremely limited information exists for short GRBs, motivates future such searches with {\it HST}, {\it JWST}, and upcoming extremely large telescopes. We come to the following conclusions.

\begin{itemize}
    \item The joint X-ray, NIR, and radio observations cannot be explained as synchrotron emission from the GRB forward shock alone.
    \item While the radio and X-ray emission can be well fit to a forward shock, this model under-predicts the observed NIR emission by factors of $\approx 5$--$10$, leaving an ``excess'' of NIR (rest-frame optical) emission.
    \item The X-ray and radio luminosity and temporal evolution of \grb\ is comparable with that of other cosmological short GRBs. However, the NIR counterpart ($\approx 10^{42}$~erg~s$^{-1}$) is sub-luminous in comparison with detected short GRB afterglows, and an order of magnitude brighter than any known kilonova or kilonova candidate.
    \item We propose that the NIR (rest-frame optical) excess emission could be a kilonova boosted by energy deposition from a stable magnetar remnant, or a radioactively-powered kilonova with modified ejecta or heating properties relative to GW170817.
    \item An alternative explanation for the broad-band emission of \grb\ is a forward shock with a relatively wide jet opening angle of $\approx 14^{\circ}$. In this model, the predicted X-ray decline rate is steeper than observed, while the early radio emission is under-predicted, the latter of which can be reconciled with the addition of a reverse shock component.
    \item \grb\ originated in a bright region of its host galaxy, at a projected offset of $\approx 1$~kpc, or $\approx 0.24r_e$, from the center (closer than 90\% of short GRBs). The host galaxy is a young ($\approx 0.53$~Gyr), modestly star-forming galaxy (SFR$\approx2.1$--4.8$M_{\odot}~{\rm yr}^{-1}$) galaxy with $M_* \approx 4.5 \times 10^{9}\,M_{\odot}$.
    \item The detection of the NIR (rest-frame optical) counterpart to \grb\ may contribute to the diversity of counterparts observed accompanying GW-detected BNS mergers. Current (upcoming) optical searches following GW events will be sensitive to such counterparts to $\approx\!160$--$250$~Mpc ($\approx 600$~Mpc). However, if the emission of \grb\ resulted from a magnetar, the fraction of BNS mergers with such high-luminosity counterparts is expected to be low.
    \item If the progenitor of \grb\ did indeed produce a stable magnetar, late-time synchrotron radio emission is predicted to become observable with the VLA on $\sim\!0.3$--$6$~year timescales, and peak at $\approx 1$--$10$~years, with the range depending on the ejecta and environmental properties.
\end{itemize}

Our work demonstrates the power of multi-epoch afterglow observations for host galaxy association and uncovering the surprising diversity of broad-band properties in short GRBs. Early radio observations of short GRB afterglows at $\lesssim1$~day are key to capturing reverse shock signatures, and to constraining the composition of their jets. On the other hand, multi-frequency observations at $1$--10~days are vital for constraining the ejecta collimation and deriving the true cosmological rate of compact object mergers in the era of Advanced LIGO. Future late-time $\gtrsim5$--10~yr, sensitive ($\approx1\,\mu$Jy) radio searches may be used to test for the presence of the radio emission from any magnetar produced in this and other short GRBs. Such observations in the SKA and ngVLA era may routinely be used to probe the parameter space of initial ejecta mass and magnetic field, thereby constraining magnetar formation and spin-down models, and yielding further insight into the GRB central engine and progenitor channels. 

\section*{Acknowledgements}
W.F. thanks her research group and collaborators for their inspiration, tenacity and strength during this time. We thank Amy Lien, Joel Leja, Stijn Wuyts, and Sarah Wellons for helpful discussions. 

The Fong Group at Northwestern acknowledges support by the National Science Foundation under grant Nos. AST-1814782 and AST-1909358.  Support for program \#15964 was provided by NASA through a grant from the Space Telescope Science Institute, which is operated by the Association of Universities for Research in Astronomy, Inc., under NASA contract NAS 5-26555. G.S. acknowledges for this work was provided by the NSF through Student Observing Support award SOSP19B-001 from the NRAO. AEN acknowledges support from the Henry Luce Foundation through a Graduate Fellowship in Physics and Astronomy. YD acknowledges support for the Illinois Space Grant Undergraduate Research Fellowship through the Illinois Space Grant Consortium by a NASA-awarded educational grant. C.D.K. acknowledges support through NASA grants in support of {\it Hubble Space Telescope} programs GO-15691 and AR-16136. J.B. is supported by the National Aeronautics and Space Administration (NASA) through the Einstein Fellowship Program, grant No. PF7-180162. MN is supported by a Royal Astronomical Society Research Fellowship.
K.D.A. is supported by NASA through the NASA Hubble Fellowship grant  \#HST-HF2-51403.001-A awarded by the Space Telescope Science Institute, which is operated by the Association of Universities for Research in Astronomy, Incorporated, under NASA contract NAS5-26555.
B.M. is supported by NASA through the NASA Hubble Fellowship grant \#HST-HF2-51412.001-A awarded by the Space Telescope Science Institute, which is operated by the Association of Universities for Research in Astronomy, Inc., for NASA, under contract NAS5-26555. A.C. and R.S. acknowledges support by the National Science Foundation under grant No. HBCU-UP AST-1831682.

This research is based on observations made with the NASA/ESA Hubble Space Telescope obtained from the Space Telescope Science Institute, which is operated by the Association of Universities for Research in Astronomy, Inc., under NASA contract NAS 5–26555. These observations are associated with program \#15329.
W. M. Keck Observatory access was supported by Northwestern University and the Center for Interdisciplinary Exploration and Research in Astrophysics (CIERA). Some of the data presented herein were obtained at the W. M. Keck Observatory (PI Blanchard; Program O287), which is operated as a scientific partnership among the California Institute of Technology, the University of California and the National Aeronautics and Space Administration. The Observatory was made possible by the generous financial support of the W. M. Keck Foundation. The authors wish to recognize and acknowledge the very significant cultural role and reverence that the summit of Maunakea has always had within the indigenous Hawaiian community.  We are most fortunate to have the opportunity to conduct observations from this mountain.
The National Radio Astronomy Observatory is a facility of the National Science Foundation operated under cooperative agreement by Associated Universities, Inc. This work is based in part on observations made with the Spitzer Space Telescope, which was operated by the Jet Propulsion Laboratory, California Institute of Technology under a contract with NASA. This work makes use of observations from the Las Cumbres Observatory global telescope network using the 1-m Sinistro instrument at the Sutherland South African Astronomical Observatory site.

This research was supported in part through the computational resources and staff contributions provided for the Quest high performance computing facility at Northwestern University which is jointly supported by the Office of the Provost, the Office for Research, and Northwestern University Information Technology. This work made use of data supplied by the UK \textit{Swift} Science Data Centre at the University of Leicester.

This work is based in part on observations made with the Spitzer Space Telescope, which was operated by the Jet Propulsion Laboratory, California Institute of Technology under a contract with NASA.
Funding for SDSS-III has been provided by the Alfred P. Sloan Foundation, the Participating Institutions, the National Science Foundation, and the U.S. Department of Energy Office of Science. The SDSS-III web site is http://www.sdss3.org/. SDSS-III is managed by the Astrophysical Research Consortium for the Participating Institutions of the SDSS-III Collaboration including the University of Arizona, the Brazilian Participation Group, Brookhaven National Laboratory, Carnegie Mellon University, University of Florida, the French Participation Group, the German Participation Group, Harvard University, the Instituto de Astrofisica de Canarias, the Michigan State/Notre Dame/JINA Participation Group, Johns Hopkins University, Lawrence Berkeley National Laboratory, Max Planck Institute for Astrophysics, Max Planck Institute for Extraterrestrial Physics, New Mexico State University, New York University, Ohio State University, Pennsylvania State University, University of Portsmouth, Princeton University, the Spanish Participation Group, University of Tokyo, University of Utah, Vanderbilt University, University of Virginia, University of Washington, and Yale University. The Pan-STARRS1 Surveys (PS1) and the PS1 public science archive have been made possible through contributions by the Institute for Astronomy, the University of Hawaii, the Pan-STARRS Project Office, the Max-Planck Society and its participating institutes, the Max Planck Institute for Astronomy, Heidelberg and the Max Planck Institute for Extraterrestrial Physics, Garching, The Johns Hopkins University, Durham University, the University of Edinburgh, the Queen's University Belfast, the Harvard-Smithsonian Center for Astrophysics, the Las Cumbres Observatory Global Telescope Network Incorporated, the National Central University of Taiwan, the Space Telescope Science Institute, the National Aeronautics and Space Administration under Grant No. NNX08AR22G issued through the Planetary Science Division of the NASA Science Mission Directorate, the National Science Foundation Grant No. AST-1238877, the University of Maryland, Eotvos Lorand University (ELTE), the Los Alamos National Laboratory, and the Gordon and Betty Moore Foundation.

\vspace{5mm}
\facilities{\textit{Swift}(XRT and UVOT), Keck:I (LRIS), Keck:II (DEIMOS), VLA, LCOGT}

\software{CASA \citep{CASA}, \texttt{Prospector} \citep{Leja_2017}, \texttt{Python-fsps} \citep{FSPS_2009, FSPS_2010}, \texttt{Dynesty} \citep{Dynesty}, \texttt{HEASoft} \citep{Blackburn1999,NASA2014}, \texttt{Xspec} \citep{arn96}}

\bibliography{refs}

\end{document}